\newcommand{\Msun}{\mbox{$M_{\odot}$}}
\newcommand{\Lsun}{\mbox{$L_{\odot}$}}

\newcommand{\Rsun}{\mbox{$R_{\odot}$}}
\newcommand{\Meff}{\mbox{$M_{\rm eff}$}}
\newcommand{\Meffp}{\mbox{$M_{\rm eff}({\rm p})$}}
\newcommand{\Meffeq}{\mbox{$M_{\rm eff}({\rm eq})$}}

\newcommand{\Teff}{\mbox{$T_{\rm eff}$}}

\newcommand{\Tmean}{\mbox{$T_{\rm mean}$}}

\newcommand{\vrot}{\mbox{$v_{\rm rot}$}}
\newcommand{\vinf}{\mbox{$v_{\infty}$}}

\newcommand{\vesc}{\mbox{$v_{\rm esc}$}}
\newcommand{\vescp}{\mbox{$v_{\rm esc}({\rm p})$}}
\newcommand{\vesceq}{\mbox{$v_{\rm esc}({\rm eq})$}}
\newcommand{\vroteq}{\mbox{$v_{\rm rot}({\rm eq})$}}
\newcommand{\vcrit}{\mbox{$v_{\rm crit}$}}
\newcommand{\geff}{\mbox{$g_{\rm eff}$}}

\newcommand{\Msunyr}{\mbox{$M_{\odot}/yr$}}
\newcommand{\kapatm}{\mbox{$\kappa_{\rm atm}$}}
\newcommand{\Ledatm}{\mbox{$L_{\rm E}^{\rm mod} $}}

\newcommand{\Gamatm}{\mbox{$\Gamma_{\rm mod} $}}
\newcommand{\Gammae}{\mbox{$\Gamma_{\rm e} $}}
\newcommand{\Gameq}{\mbox{$\Gamma_{\rm e}({\rm eq})$}}
\newcommand{\Gamp}{\mbox{$\Gamma_{\rm e}({\rm p})$}}
\newcommand{\gn}{\mbox{$g_{\rm N} $}}
\newcommand{\gN}{\mbox{$g_{\rm N} $}}
\newcommand{\gE}{\mbox{$g_{\rm E}$}}
\newcommand{\gedd}{\mbox{$g_{\rm E}$}}

\newcommand{\grad}{\mbox{$g_{\rm rad}$}}

\newcommand{\Rtheta}{\mbox{$R(\Theta)$}}
\newcommand{\Rp}{\mbox{$R_{\rm p}$}}
\newcommand{\Req}{\mbox{$R_{\rm eq}$}}
\newcommand{\sigmae}{\mbox{$\sigma_{\rm e}$}}
\newcommand{\sigmaeUF}{\mbox{$\sigma_{\rm eUF}$}}

\newcommand{\Omegacrit}{\mbox{$\Omega_{\rm crit}$}}
\newcommand{\Frad}{\mbox{$F_{\rm rad}$}}

\newcommand{\Teq}{\mbox{$T_{\rm eq}$}}
\newcommand{\Tp}{\mbox{$T_{\rm p}$}}

\newcommand{\Lp}{\mbox{$L_{\rm p}$}}
\newcommand{\Ltot}{\mbox{$L_{\rm tot}$}}

\newcommand{\kapmax}{\mbox{$\kappa^{\rm max} $}}
\newcommand{\kapmaxUF}{\mbox{$\kappa^{\rm max}_{\rm UF} $}}
\newcommand{\kztx}{\mbox{$k(Z,T,X_{\rm H})$}}
\newcommand{\fT}{\mbox{$f(T_{\rm eff})$}}
\newcommand{\mue}{\mbox{$\mu_{\rm e}$}}

\newcommand{\Mtams}{\mbox{$M_{\rm TAMS}$}}
\newcommand{\Ltams}{\mbox{$L_{\rm TAMS}$}}
\newcommand{\Rtams}{\mbox{$R_{\rm TAMS}$}}
\newcommand{\Xhtams}{\mbox{$X_{\rm TAMS}$}}
\newcommand{\Ymax}{\mbox{$Y_{\rm max}$}}
\newcommand{\Xh}{\mbox{$X_{\rm H}$}}
\newcommand{\Mi}{\mbox{$M_{\rm i}$}}

\newcommand{\omi}{\mbox{$\omega_{\rm i}$}}

\newcommand{\LMcrit}{\mbox{$(L/M)_{\rm crit} $}} 
\newcommand{\cross}{\mbox{${\rm cm}^2\ {\rm g}^{-1} $}}

\documentclass{aastex62}
\usepackage{graphicx,amsmath}

\begin{document}

\title{The Effect of Rotation on Triggering S Doradus Instabilities in Luminous Blue Variables}

\author{Emily M. Levesque}
\affil{Department of Astronomy, Box 351580, University of Washington, Seattle, WA 98195, USA}
\email{emsque@uw.edu}

\author{Henny J. G. L. M. Lamers}
\affil{Astronomical Institute Anton Pannekoek, University of Amsterdam,  Science Park 904, NL-1098XH, Amsterdam, The Netherlands}
\email{h.j.g.l.m.lamers@uu.nl}

\author{Alex de Koter}
\affil{Astronomical Institute Anton Pannekoek, University of Amsterdam,  Science Park 904, NL-1098XH, Amsterdam, The Netherlands}
\email{A.deKoter@uva.nl}

\begin{abstract}
Luminous blue variables are an intermediate stage in the evolution of high-mass stars characterized by extreme mass loss and substantial variability. The stars show large irregular episodic variations on timescales of years to decades in these stars' effective temperatures (called ``S Dor variations"). Observations show that these variations are triggered when the stars are in a well-defined strip in the HRD that corresponds to the Modified Eddington Limit, where the atmospheric radiation pressure almost balances gravity. In this work we consider the role that rotation plays in the instability that leads to the triggering of S Dor variations in luminous post-main sequence LBVs. We adopt the existing instability criterion 
that the effective surface gravity is reduced to 10\% of the Newtonian gravity due to radiation pressure in the atmosphere of non-rotating stars. We then specifically describe how rotation impacts this instability. By carrying out numerical simulations of model LBVs at both solar and sub-solar metallicities, we confirm that most LBVs should be unstable at both the equator and the poles, and that rotation exacerbates this effect; some models also produce enhanced mass loss at the pole or equator. Our numerical models also predict dense equatorial disks or rings and high-velocity bipolar outflows, in agreement with existing observations of LBV circumstellar nebulae.  \\
\end{abstract}

\section{Introduction}
Luminous blue variables (LBVs) are an intermediate stage in the evolution of high mass ($\gtrsim30M_{\odot}$) stars. They represent a crucial evolutionary stage for these stars, characterized by extreme mass loss and substantial variability. Despite their importance there are a number of open questions regarding LBVs.

Telltale physical signatures of LBVs include their extremely high luminosities (log($L$/\Lsun)$\gtrsim$5.4) and large irregular episodic variations in their temperatures on timescales of years to decades. The latter are referred to as ``S Dor variations". During S Dor variations LBVs vary between a ``hot" state and a ``cool" state (hitting a fairly narrow minimum $T_{\rm eff}$ range of 7500-9000 K; Figure 1) while maintaining a nearly constant luminosity.  During the ``hot state" LBVs occupy a narrow well-defined strip in the HRD, sometimes referred to as the S Dor instability strip, that runs from about $(\log(L/\Lsun), \log(\Teff)) \simeq (6.2,4.5)$ to (5.4, 4.1) (e.g. \citealt{Wolf89}, \citealt{HumphreysDvidson94},\citealt{Dekoter96} \citealt{SmithVinkdeKoter04}, \citealt{WeisBomans20}, \citealt{Davidson20}, and Figure 1). The transition between the two states manifests observationally as long-term photometric variability due to the star's changing bolometric correction as a function of $T_{\rm eff}$.

\cite{Davidson20} has pointed out the potential difference in evolutionary phase between the more luminous LBVs (log $L>5.7$) and the lower luminosity LBVs (log $L<5.7$) (see also \citealt{Wolf89}). While both experience similar S Dor variations, some of the more luminous LBVs have also experienced giant eruptions such as those associated with $\eta$ Car and P Cyg. The explanation of these giant eruptions has been the topic of several recent theoretical studies. (e.g. \citealt{Owocki17}, \citealt{Owocki19}, \citealt {OwockiShaviv16}, and \citealt{Davidson20}.
\citet{Jiang18} also studied the structure and stability of LBVs by means of 3D hydrodynamic simulations
and suggested an explanation for fast low luminosity variations on a timescale of days.

In this paper we concentrate on the S Dor variations that occur on timescales of years. The large variations in \Teff\ at almost constant luminosity during S Dor variations indicate drastic variations in the effective radius of the stars, on timescales of years to decades. These variations can be as large as a factor of $\sim2$ (R110, $\Teff =$ 7600-10300 K) to  $\sim10$ (AG Car, $\Teff =$ 9000-30000 K) with the more luminous LBVs showing the largest variations. These changes are not primarily due to variations in the mass-loss rate and the associated changes in the effective radius in the wind, because the observed changes in the mass-loss rate are not large enough \citep{Dekoter96, VinkDekoter02, SmithVinkdeKoter04}. They must therefore be due to large radius variations in the subphotospheric layers.

Several mechanisms have been suggested to explain these episodic S Dor variations (see reviews by \citealt{Vink12} and \citealt{Davidson20}). They are all based on dynamical instabilities of the sub-photospheric layers and a violent reaction of the atmosphere which is at the limit of stability against radiation pressure. 
 The luminosity of LBVs is below the classical Eddington limit for radiation pressure by electron scattering, but the sub-photospheric envelope opacity is larger than the electron scattering opacity. If the star is close to the luminosity limit for electron scattering, any additional opacity in the envelope will bring the atmosphere close to the stability limit where $\grad \simeq \gn$. This is called the ``Modified Eddington Limit" (MEL; \citealt{app86, Lamers86, Davidson87, Davidson20}). A small disturbance in the subphotospheric layers may then disrupt the delicate stability, and trigger the observed S Dor variations. In this respect, the iron opacity peak around a temperature of $\sim$200000 K appears to play a key role.

 This suggestion is supported by the model of \cite{grassitelli21} that roughly explains both the timescale of the S Dor variations and the range of the observed luminosity changes. They calculated a time-dependent stellar evolution model for a non-rotating 100\Msun\ star near the Eddington limit,  
with sonic-point boundary conditions
for wind models with mass-loss rates from \cite{Vink01} and with a beta-velocity law with exponent unity \citep{GraefenerHamann08}
and \vinf\ = 2.6, or 1.3 times \vesc\ at the 
hot and cool side of the bistability temperature of \Teff\ = 21000 K \citep{LamersSnowLindholm95}. The feedback of the mass loss on the envelope structure was taken into account. When the mass-loss rate increases above a critical limit due to the bistability jump, the opacity decreases and the extended envelope adjusts itself (shrinks) on the thermal timescale of the inflated envelope (years). This, in turn,  leads to mass-loss variations that reverse the initial variations, resulting in a cycle that lacks a stationary equilibrium.

Studies of the MEL by \cite{UF98} (hereafter UF98) have shown that the location in the HRD of LBVs during visual minimum (often called ``quiesence") coincides approximately with the location where the radiative accelation in their atmospheres reaches about 90\% of their gravity if rotation is neglected. This strongly suggests that the S Dor variations are triggered when the effective gravity (i.e. corrected for radiation pressure in the atmosphere) is very low. A low
effective gravity implies a high mass-loss rate which is an ingredient in the instability model of \cite{grassitelli21}. 

Observations of high rotation rates in LBVs (ranging from tens to hundreds of km s$^{-1}$, e.g. \citealt{Davies05,Groh06}) demonstrate that rotation effects must also play a significant role in at least some LBVs. 
Indeed, \cite{langer97} already suggested that rapid rotation is responsible for the occurrence of both giant eruptions and, possibly, the more typical S Dor variations in LBVs.
Rotation in massive stars can significantly impact their evolution and physical properties. At the surface of a star, rapid rotation leads to mechanical changes (i.e., larger equatorial radii and oblateness of gravitational equipotential surfaces) and associated photospheric effects.  The most notable of the latter is a decreased $T_{\rm eff}$ at the equator as compared to the poles, a consequence of the decrease in effective gravity at the equator and the local radiative flux being proportional to the local effective gravity \citep{VonZeipel24,Eddington26}.

In this paper we consider the role of rotation in the location of the MEL in the H-R diagram. This location thus indicates the stellar parameters where subphotospheric instabilities could trigger the observed S Dor variations of LBVs. Note that we do not study S Dor variations themselves, but rather the effects of rotation on the atmospheric conditions that could trigger these variations. Nor do we suggest that fast rotation is {\it required} for triggering the S Dor variations; rather, it helps to reduce the effective gravity and increase the mass-loss rate, which is a requirement for triggering the S Dor variations \citep{grassitelli21}.

Our work concentrates on stars with log $L\gtrsim5.5$ (corresponding to stars with initial masses of $\ge$32~M$_{\odot}$, e.g. \citealt{Ekstrom12,Georgy13}), as this agrees well with the lowest-luminosity stars that are {\it confirmed} LBVs with well-studied S Dor variations (see Table 1 and Fig. 1). The minimum luminosity for the LBVs we consider here is close to the maximum luminosity  $\log(L/\Lsun)=5.5$ of red supergiants (RSGs). \cite{DornWallenstein23}, in their study of 5000 cool supergiants in the LMC based on GAIA-XP spectra, find no RSGs above $\log(L/\Lsun) > 5.5$, and \cite{Martin23} identify only two RSGs with luminosity between $5.5 < \log(L/\Lsun) < 5.7$ (though it is important to note that these same studies do identify several warmer evolved stars in the LMC with $3.7 < \log(T_{\rm eff}) < 4.2$ and $5.5 < \log(L/\Lsun) < 5.7$, including yellow hypergiants; see \citealt{DornWallenstein20,DornWallenstein23,Martin23}). \cite{Massey23} presents a detailed discussion on the upper luminosity limit of RSGs, which they find to be $\log(L/\Lsun) \sim 5.5$. We limit our focus here to stars with initial mass of $M \ge 32\Msun$ that may become LBVs shortly after leaving the main sequence, and do not include (presumably-lower-luminosity) stars that may be in the post-RSG phase.

In Section 2 we describe the properties of LBVs, including their location and variations in the HRD and their observed rotation rates. In Section 3 we describe the dependence of the MEL on stellar parameters 
and abundance for non-rotating stars. In Section 4 we express the role of rotation on the MEL of rotationally-distorted stars, and we derive a criterium for the triggering of the S Dor variations 
at the poles or at the equator.
Section 5 uses these results to describe how this instability is triggered in rotating model LBVs. Section 6 then applies these results in numerical simulations of LBVs and considers observable signatures of this mechanism, while Section 7 compares the results of our numerical simulations to existing observations. Finally, we discuss the implications of these results and the potential for future work in Section 8.


 \section{Properties of LBVs} \label{sec:variations}
 \subsection{LBVs on the H-R Diagram} \label{sec:LBVproperties}
 
 The primary identifying characteristic of confirmed LBVs is the S Dor spectroscopic and photometric variability described above, shifting between a hot state (with associated small radii, $\sim$50$-$150$R_{\odot}$) and a cool state (with larger radii, $\sim$200$-$500$R_{\odot}$). These variations are irregular and occur on a timescale of years to decades (see e.g. lightcurves collected by \citealt{Spoon94} and \citealt{Stahl01}.) 

A sample of Galactic and LMC LBVs with well-known physical properties during both the cool and hot phases is listed in Table 1, in order of decreasing luminosity. It should be noted that the small sample listed in Table 1 is comprised of {\it confirmed} LBVs with established S Dor variability (including S Dor itself) and well-defined physical properties in both their hot and cool states. This excludes, for example, the well-known but unusual stars $\eta$ Carinae and P Cygni that do not show S Dor variability. This sample also excludes candidate LBVs whose S Dor variability has not yet been observationally confirmed and those that lack robust physical properties during both the hot and cool states (see Table 1).

We point out that the area in the HRD where the LBVs reside also contains many ``normal" massive stars that do not show S Dor variations. We discuss this in \S8, after we have shown that the triggering of the S Dor variations is extremely sensitive to the $L/M$ ratio and the rotation rate.

\begin{deluxetable*}{l l l l l l}
 \tablewidth{0pt}
 \tabletypesize{\footnotesize}
 \tablecolumns{6}
 \tablecaption{Physical Properties of LBVs With Well-Observed S Dor Variations}
 \tablehead{
     \colhead{Galaxy} & \colhead{Star} & \colhead{log($L$/$L_{\odot}$)} & \colhead{$T_{\rm min}$ (K)} & \colhead{$T_{\rm max}$ (K)} & \colhead{Ref.}\tablenotemark{a} 
     } 
    \startdata
    MW  &    AG Car &  6.10 &  30000 &  9000 & 1 \\ 
    LMC  &     R 127 &  6.06 &  27000 &  8500 & 1,2 \\ 
    M33  &     Var B & 6.05 &  20000 &  9000 & 1,3 \\ 
    MW  &    W243  &  5.86 &  18500 &  8500    & 4,5 \\ 
    M33  &    Var C &  5.82 &  21600 &  7900    & 1,6,7 \\ 
    LMC  &     S Dor &  5.82 &  20000 &  9000    & 1,8 \\ 
    LMC  &     R  71 &  5.78 &  17200 &  9000    & 1,9 \\ 
    MW  &     HR Car&  5.70 &  17900 &  8000    &10,11,12 \\ 
    M31  &     LAMOSTJ0037+4016 &5.65 &15200 &9400 &6 \\ 
    SMC  &     R  40 &  5.50 &  12000 &  8700   & 1 \\ 
    MW  &  HD160529 &  5.46 &  11000 &  8000   & 1 \\ 
    LMC  &     R 110 &  5.46 &  10300 &  7600   & 1 \\ 
    \enddata
 \tablenote{References: 1=\citet{HumphreysDvidson94}, 2=\citet{Crowther97}, 3=\citet{Szeifert96}, 4=\citet{ClarkNegueruela04}, 5=\citet{Ritchie09}, 6=\citet{Huang19}, 7=\citet{Burggraf15}  8=\citet{Lamers95}, 9=\citet{Mehner17}, 10=\citet{Groh09}, 11=\citet{vanGenderen01}, 12=\citet{Mehner21}}
 \end{deluxetable*}

\begin{figure}
\centering
\includegraphics[width=14cm]{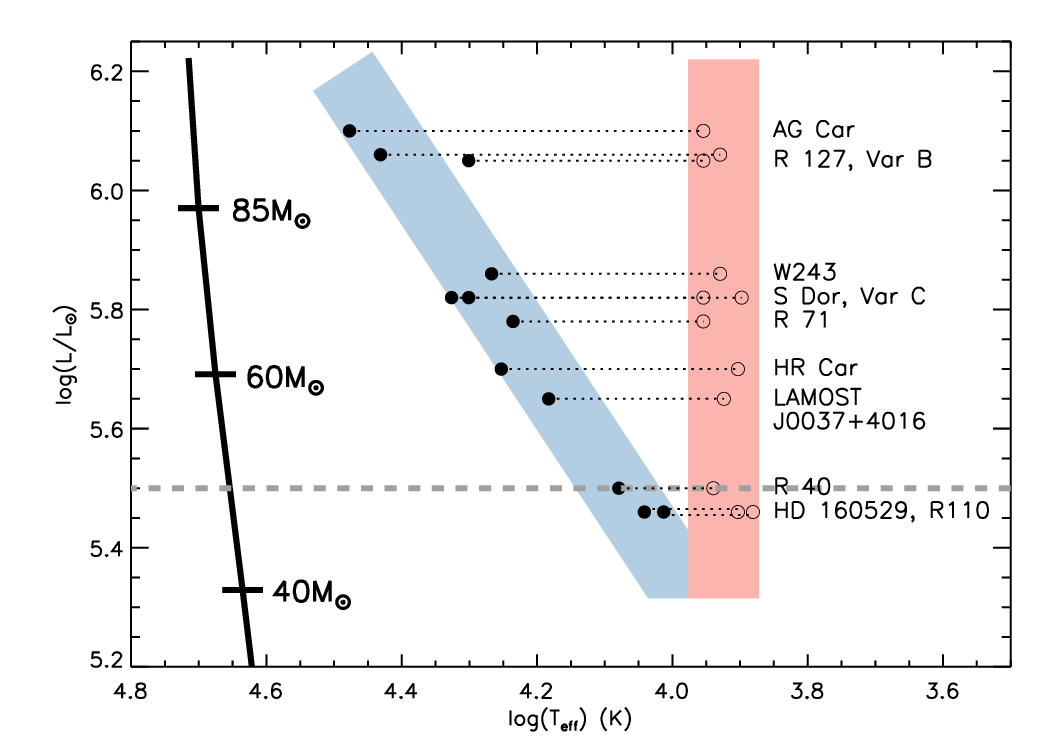}  
\caption{
  The location on the HRD of LBVs from Table 1 with well-determined physical properties during both the ``hot" and ``cool" phases of their S Dor variability. Filled circles indicate the stars'
  location during \Teff\ maximum (optical minimum) and are connected with dotted lines to open circles representing the same stars during \Teff\ minimum (optical maximum). LBVs in their hot state occupy a narrow strip of \Teff\ increasing with $L$, with a slope of $ {\rm d log}(L)/ {\rm d log}(\Teff) \sim 1.63$ (blue band). On the other hand, LBVs in their cool state lie along a vertical band of $\Teff \simeq 9000$ K (red band); this band is empirically determined, but is also understood as a consequence of the MEL. For reference, the predicted zero age main sequence for stars with solar metallicity is shown by a full black line with values of initial masses indicated. The gray dashed line at $\log(L/\Lsun)=5.5$ corresponds to the adopted upper luminosity limit of RSGs (see Introduction).
}
\label{fig:fig1}
\end{figure}

\subsection{Rotation in LBVs}
The Geneva rotating evolutionary tracks by \citet{Ekstrom12}, \citet{Georgy13}, and \citet{Groh19} model the evolution of stars with an initial rotation rate of $\omega$=0.4, or 40\% of the star's critical or ``break-up" angular velocity, and illustrate a number of key effects that rotation can have on massive stellar evolution. These include extended main sequence lifetimes, enhanced surface mixing, stronger mass loss, and changes in the terminal state of supernova progenitors. However, it is worth noting that the adopted initial rotation rate of $\omega$=0.4 for these Geneva models, chosen based on the peak of the velocity distribution observed in young B stars \citep{Huang10}, is now generally considered high compared to the normal population of massive main-sequence stars (see, for example, discussions in \citealt{Schneider14,Holgado22,Keszthelyi22}, and \citealt{Bodensteiner23}).

However, we do see evidence of extremely high rotation ($\omega \ge 0.8$) in at least some LBVs, such as AG Car ($\omega \sim 0.86$, \citealt{Groh06, Groh09}) and HR Car ($\omega \sim 0.88$, \citealt{Groh09}). Explaining why LBV rotation rates diverge from the typical massive star population is beyond the scope of this work, but potential avenues for LBV spin-up include preferentially-polar mass loss (e.g. \citealt{Maeder02}) or angular momentum gained as a result of recent mergers, a progenitor scenario that has been proposed as one potential avenue for forming LBVs (\citealt{Smith15}; however, note that this scenario has been questioned by subsequent work including \citealt{Humphreys16}, \citealt{Davidson16}, and \citealt{Aadland18}). Regardless, it is clear that a comprehensive model of instabilities in the atmospheres of LBVs must include the effects of rotation.

 
\section{The Modified Eddington Limit and the trigger of the S Dor variation} \label{sec:AEL}

The S Dor variations of LBVs are most likely triggered by 
 a subphotospheric instability that brings the star out of dynamical equilibrium. This happens in the evolutionary phase when the star approaches the MEL, where the radiation pressure force in the atmosphere almost equals the gravity.
 
 \cite{SmithVinkdeKoter04} have
  shown that the mass-loss rates increase by about a factor of 3 when $\Gamatm \equiv g_{\rm rad}/g_{\rm N}$ increases from
  about 0.4 to 0.8.
  The mass loss is expected to
  increases even more, possibly as much as a factor $>10$, if $\Gamatm$ increases to  $\simeq 0.9$ \citep{Maeder09,Vink11,VinkGraefner12}.
 
 
 \subsection{The MEL of non-rotating stars}
 \label{sec:AEL-rot}
 
  After their main sequence phase, during H-shell fusion, massive luminous stars evolve to the right in the HRD at almost constant luminosity. As \Teff\ decreases, the opacity in the atmosphere
  increases due to the bound-bound transitions of singly- and doubly-ionized metals, mainly Fe,
  reaching a peak around $\Teff \simeq 10000$ to 15000 K (e.g. \citealt{app86,Appenzeller89,Davidson87,Lamers86,LamersNoordhoek93,HumphreysDvidson94,Asplund98}).
  
  The increase in atmospheric opacity with decreasing \Teff\ implies an increase in
  radiation pressure and a decrease in the MEL (\Ledatm), defined by

  \begin{equation} \label{eq:latm}
    \Ledatm \equiv 4 \pi c GM/\kapatm = (\sigmae / \kapatm) L_{\rm E} ,
  \end{equation}
  where $L_{\rm E}$ is the classical Eddington limit for electron scattering, $\sigmae = \sigma_{\rm T}/(\mue m_{\rm H}) \simeq 0.200(1+\Xh)$ \cross\ is the absorption coefficient for electron scattering,  and $\kapatm > \sigmae$.
    In this expression $X_H$ is the photospheric H abundance and $\kapatm$
  is the maximum Rosseland mean opacity in the photosphere between $10^{-2} < \tau < 10^{3}$ \citep{Davidson87,LamersFitzpatrick88,UF98}.
  
  The increasing influence of radiation pressure in the atmosphere during the expansion can be expressed in terms of the Modified Eddington factor \Gamatm,

  \begin{equation} \label{eq:gamatm}
    \Gamatm \equiv g_{\rm rad} / \gn = \kapatm L / 4 \pi c G M
  \end{equation}
  where $\gn =GM/R^2$ is the Newtonian gravity.
  
\begin{figure}
\centering
\includegraphics[width=14cm]{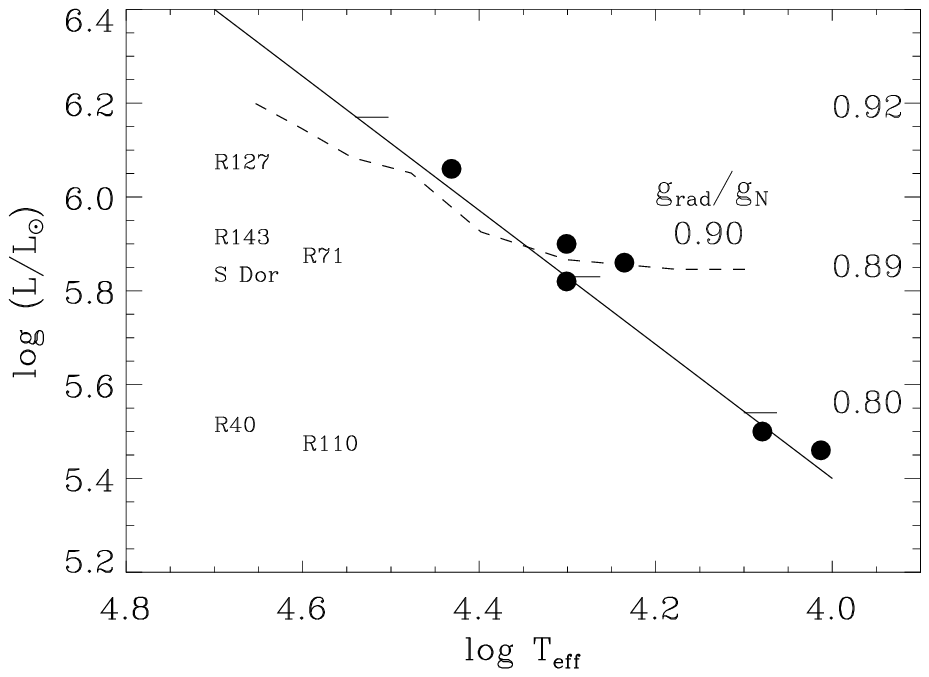}  
\caption{
  The location in the HRD of LMC and SMC LBVs in their hot phase is compared with their predicted location for various values of $\grad/\gN$. The solid line is the LBV instability strip. The dashed line is the location where $\grad/\gN=0.90$. The location where several values of this parameter (0.80, 0.89, and 0.92) are reached on the
  LBV-strip are indicated with short horizontal lines.
}
\label{fig:fig1A}
\end{figure}

  UF98 have argued on the basis of LTE model atmospheres that a massive  star becomes unstable if 
  $\grad / \gn \ge 0.9$ (i.e. the
  effective gravity is reduced to only 10\% of the Newtonian gravity). They show that for non-rotating stars the predicted location of massive post-MS stars
  in the HRD agrees with the Humphreys-Davidson limit, which in turn agrees approximately with the S Dor instability strip that corresponds to the locations of LBVs during their hot phase.
  In Figure \ref{fig:fig1A}  we compare the location in the HRD of the LMC and SMC LBVs with the predicted location for different values of $\grad/\gn$. This comparison is based on the calculations of the maximum photospheric opacity at $10^{-2} < \tau < 10^{3}$ for atmospheric LTE models of $Z=0.002$ and on the 
  predicted  $L/M$-ratios from the evolutionary models of initially rotating stars with $Z=0.002$ \citep{Georgy13}. Despite the simplicity of the predictions (in particular the LTE atmospheres used) the results confirm the suggestion that the LBV strip marks the location where $\Gamatm \simeq 0.8$ to 0.9 (see also, for example, \citealt{Davidson87,HumphreysDvidson94,Humphreys16}).

 This suggests that the S Dor instability of LBVs is triggered during an evolutionary phase when
  
  \begin{equation} \label{eq:grad}
      \grad \equiv \frac{\kapmax \sigma \Teff^4}{c} = \Ymax \frac{GM}{R^2}.\footnote{In this expression the mass $M$ should be used instead of 
  $\Meff=M(1-\Gammae)$ with the Eddington factor $\Gammae = \sigmae~ \sigma \Teff^4 / c$ correcting for radiation pressure by electron scattering,
   because the model atmospheres used by UF98 are characterized by $GM/R^2$ and not by $G \Meff/R^2$. The effect of electron scattering is included in the calculation of \kapmax.}
  \end{equation}
    where $\Ymax$ is the adopted maximum value of $\Gamma_{\rm atm}$. In the rest of this study we will adopt $\Ymax =0.9$ as the characteristic value for the LBV-strip.

    Substitution of 
  $\sigma \Teff^4 = L / 4 \pi R^2$ results in a stability  criterium: 
  
  \begin{equation} \label{eq:LMcrit}
    \left( \frac{L/\Lsun}{M/\Msun}\right)_{\rm crit} = 
     \frac{\Ymax \times 4 \pi G c }{\kapmax} \cdot \frac{\Msun}{\Lsun} =
     \frac{1.17 \times 10^4}{\kapmax}.
  \end{equation}
  If the ratio $L/M$ exceeds this critical value, the atmosphere of the star will become unstable  and the S Dor variations may be triggered (a slightly smaller adopted value of \Ymax\ would correspond to an instability at a slightly lower
    value of $L/M$.)  
  
  
  \subsection{The value of \kapmax} \label{sec:kapmax}
  
  UF98 have calculated the Rosseland mean values of \kapmax\ in a large series of atmospheric models with different values of $\Teff$ and $\log(g)$.
  They used the ATLAS9 models from \cite{Kurucz95} with a turbulent velocity of 8 km s$^{-1}$, metallicities of $Z=0.02$ and $Z=0.002$, and $\Xh = 0.74$. They expressed the result in terms of 
  
   \begin{equation} \label{eq:gedd/gn}
    \gedd / \gn = \fT
  \end{equation} 
 where $\gedd = \sigmaeUF \times \sigma \Teff^4/c$ with 
    $\sigmaeUF =0.348$ cm$^2$ g$^{-1}$ for $\Xh=0.74$. Combining this with  
  Eq. \ref{eq:grad} results in a relation between \fT\ and the UF98 determination of \kapmax\ (hereafter \kapmaxUF) at the stability limit:
  
  \begin{equation} \label{eq:kapmaxf}
   \kapmaxUF(\Teff) = 0.9 \sigmae_{\rm UF}/ \fT = 0.313 / \fT ~~~~ {\rm cm}^2 {\rm g}^{-1}
  \end{equation}

 The top panel of Figure 3 shows the function $\fT$ in the temperature range of $10,000 <\Teff<60,000$ 
  for $Z=0.02$ and $Z=0.002$. 
  The lower panel shows the values of \kapmaxUF\ as function of \Teff. We see that \kapmaxUF\ increases towards lower \Teff,  reaches a maximum at around 12,000 K, and
  declines steeply at $\Teff < 11,000$ K. 
  
\begin{figure} [ht!] 
\includegraphics[width=16cm]{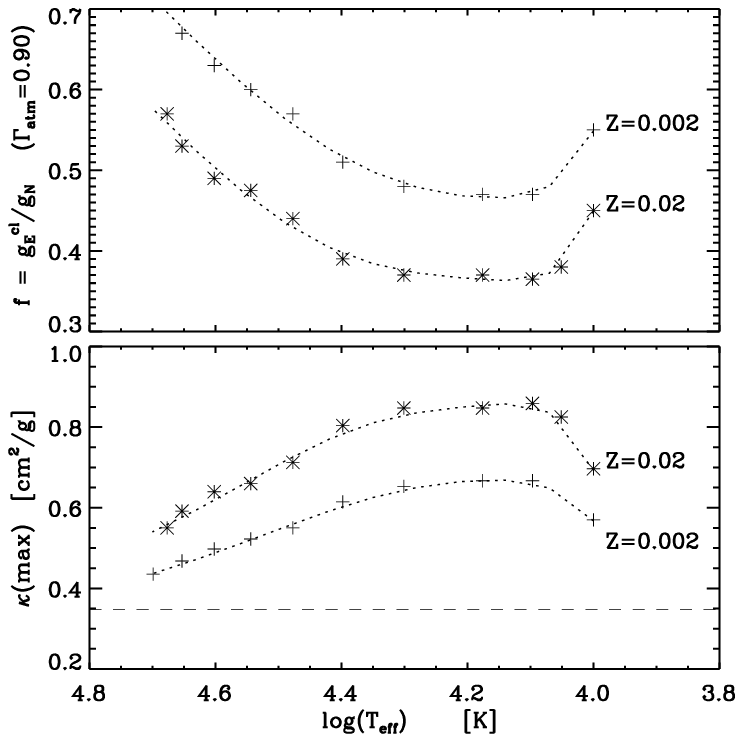}  
\caption{Upper panel: The function $\fT = \gedd / \gn$ for
  the atmospheric models of $Z=0.02$ and $0.002$ and $X=0.74$ that reach $\Gamatm=0.90$ (from UF98). Lower panel: The maximum values of
  the atmospheric absorption coefficient in the optical depth range of $10^{-2} < \tau < 10^3$ in atmospheric models of $Z=0.02$ and $0.002$ and $\Xh=0.74$ that reach $\Gamatm=0.90$, derived by UF98.
  The dashed line shows the electron scattering coefficient $\sigmae=0.348$~\cross\ for $\Xh=0.74$. 
 The dotted lines are power law fits, given in Eq. A1 of the Appendix.} 
\label{fig:fT}
\end{figure}

\begin{figure} [ht!] 
\centering
\includegraphics[width=14cm]{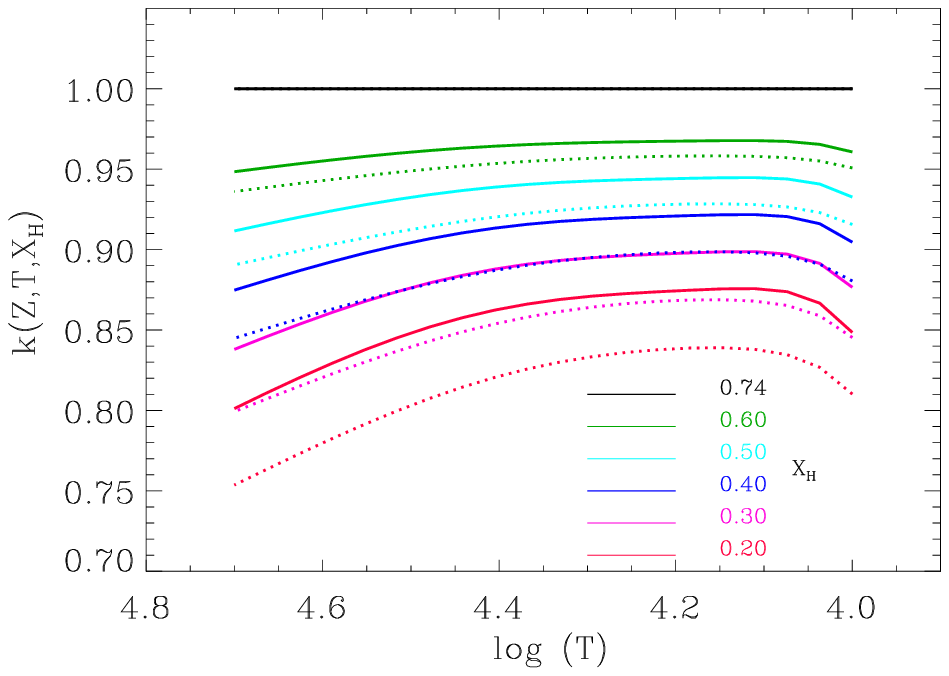}  
\caption{The function $\kztx$ for $Z=0.02$ (full lines) and
$Z=0.002$ (dotted lines) for various values of $\Xh$.
} 
\label{fig:k}
\end{figure}

The value of \kapmax\ depends strongly on the metallicity $Z$
because the primary opacity source in this regime is line and continuum absorption by metals.
The dependence of \kapmax\ on $X_H$ is less severe because H is mainly ionized in the atmospheres of the stars considered here; it mainly contributes to \kapmax\ by electron scattering, which is proportional to $(1+\Xh)$.  
In this study we will use the values derived by UF98 for $Z=0.02$ and $0.002$
and $\Xh=0.74$, with the contribution by electron scattering adjusted to the actual value of \Xh :

\begin{equation}  \label{eq:kapmaxnew}
\begin{split}
    \kapmax(\Teff) & =  \kapmaxUF(\Teff) -0.348 + \sigmae(\Xh) \\  
    & =\kapmaxUF(\Teff) + 0.200( \Xh-0.74)
\end{split}
\end{equation}
in \cross. We define the ratio

\begin{equation} \label{eq:k}
    k(Z,\Teff,\Xh) \equiv \kapmax(Z,\Teff,\Xh)/\kapmaxUF(Z, \Teff,0.74)
\end{equation}
with $Z=0.02$ or $Z=0.002$.
The function $k$ is plotted in figure \ref{fig:k}
for $Z=0.02$ and $0.002$ and for several values of $\Xh$.
The value of $k$ is always close to 1. 
The minimum values of $k$ in the range of $10,000 < \Teff < 50,000$ K
and $0.20 < \Xh < 0.74$ are 0.80 and 0.75 for $Z=0.02$ and 0.002 respectively, reached at $T=50,000$K.

  \section{Triggering the instability of rotating stars} \label{sec:rotation}
  
  Following the arguments given above for non-rotating stars we assume that in rotating post-MS LBVs the instability is also triggered when the
  effective gravity is as low as 10$\%$ of the gravity,
  i.e. $\Gamatm = 0.9$.  As the stars expand during the post-main sequence phase, the gravity decreases due to the increase in radius (roughly as $R^{-2}$), while at the same time \Teff\ decreases (as $\Teff \propto R^{-1/2}$) and $\kapatm$ increases.  The triggering, which depends on the ratio $\grad / \gn $,  might occur either at the equator or the poles of the star: rotation reduces the effective gravity at the equator, but the radiative flux of rotating stars is higher at the poles than at the equator. We consider both possibilities below.
  
  \subsection{Models of rotating stars} \label{sec:rotmodels}
  
  For a rapidly rotating star the effective gravity at stellar co-latitude $\Theta$
  (with $\Theta=0$ and $\pi$ at the poles) is 
  
  \begin{equation}  \label{eq:geff}
  \begin{split}
   \geff(\Theta) = & \Bigl[ \Bigl( \frac{ G \Meff} {R^2(\Theta)} \\ & + \Omega^2 R(\Theta)sin^2(\Theta) \Bigr)^2 \\ & + \Omega^4 R^2(\Theta) sin^2(\Theta) cos^2(\Theta) \Bigr]^{1/2}
   \end{split}
  \end{equation} 
  where $\Omega$ is the angular velocity. The first term in Eq. \ref{eq:geff} is the Newtonian gravity, corrected for the 
  atmospheric radiation pressure; the second term is the centrifugal force in the plane perpendicular to the rotation axis; and the third term is the correction for the fact that the local surface at co-latitude $\Theta$  is tilted compared to the radial vector (see Eq. 2.12 of \citealt{Maeder09}). For simplicity we have assumed that $\Omega$ is independent of latitude. We thus ignore the uncertain effects of differential rotation due to, for example, magnetic fields.

  The radius $R(\Theta)$ is the solution of the cubic equation

  \begin{equation} \label{eq:rtheta}
    \Rtheta^3~-~\frac{2G \Meff}{R(0) \Omega^2 sin^2 \Theta}\Rtheta~+~\frac{2G \Meff}{\Omega^2 sin^2 \Theta} ~=~0
  \end{equation}
  where $\Meff = M (1 - \Gamma_{\rm e})$, i.e.\ the gravity is corrected for radiation
  pressure by electron scattering. We thus assume that the shape of the star is only affected by the sub-photospheric layers where $\kappa = \sigma_{\rm e}$ and that the thickness of the photosphere
  has a negligible effect on the shape of \Rtheta.
  
  The rotation rate $\Omega$ can be expressed as a fraction of the critical
  rotation rate,  $ \omega \equiv \Omega / \Omegacrit$,  where
  
  \begin{equation}  \label{eq:omegacrit}
  \Omega_{\rm crit} = \sqrt{ G \Meff /\Req^3 }
  \end{equation}
  and \Req\ is the equatorial radius.

\begin{figure} 
  \includegraphics[width=16cm]{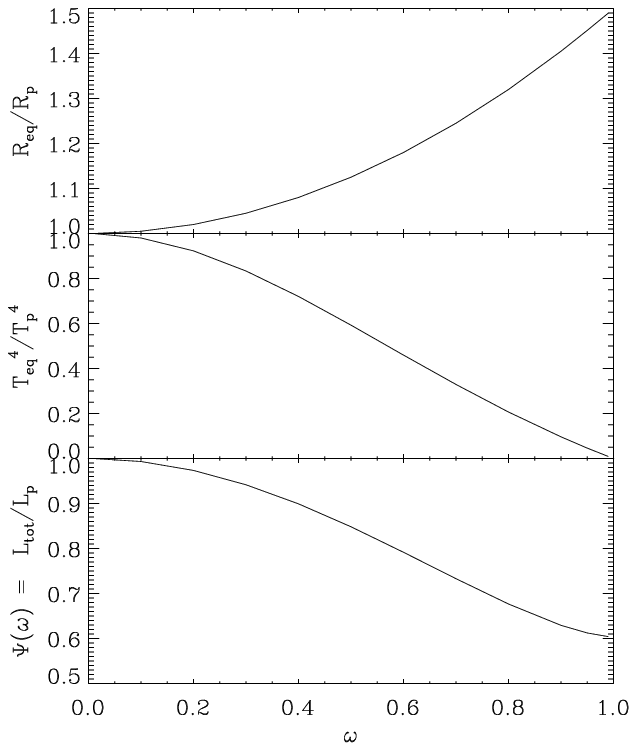}  
 \caption{Properties of rotating stars as a function of $\omega$. 
 Top:
 the ratio $\Req / \Rp$ as a function of $\omega$. Middle: the ratio $\Teq^4/ \Tp^4$ between the radiative fluxes
 at the equator and the pole.
 Bottom: The function $\Psi (\omega)= \Ltot / \Lp$.}
 \label{fig:equator-pole}
\end{figure}

  For our study of the effect of rotation on the MEL, we compare the properties of a rotating star at the pole and at the equator. 
  Substitution of $\Omega= \omega \Omegacrit$ into Eq. \ref{eq:rtheta} results in the ratio
  
  \begin{equation} \label{eq:ReRp}
   \Req/\Rp = 1 + 0.5\omega^2
   \end{equation}
   The ratio  $\Req / \Rp$  is shown in the upper panel of Figure \ref{fig:equator-pole}. 
   The figure shows that at critical rotation the 
   ratio $\Req / \Rp$ reaches a maximum of $3/2$ at $\omega=1$.

 Substitution of Eqs. \ref{eq:omegacrit} and \ref{eq:ReRp} into Eq. \ref{eq:rtheta} results in an alternative expression for the shape of the star as a function of $\Theta$ and $\omega$
 
 \begin{equation} \label{eq:rthetab}
     A ~sin^2(\Theta)~ x^3~-x~+1~=~0
 \end{equation}
 where  $A \equiv 0.5\omega^2/(1 + 0.5 \omega^2)^3$, with $0 \leq A \leq 0.1481$ and $x \equiv R(\Theta)/\Rp$ with $1 \leq x \leq 1.5$. The resulting shape can be approximated to a high accuracy as
 
 \begin{equation} \label{eq:shapefit}
     \Rtheta / \Rp = x = h(A~sin^2(\Theta))
 \end{equation}
 with $h$ given by  Eq. \ref{eq:RthetaRp} in the Appendix.

 For oblate rotating stars the local radiative flux at the
 stellar surface scales with $\geff(\Theta)$ \citep{VonZeipel24, Eddington26}, which implies that $\Frad(\Theta) \propto \Teff^4(\Theta)$ decreases from 
 the pole to the equator with $\geff(\Theta)$ described by Eq. \ref{eq:geff}.
  
\begin{figure} 
  \center
  \includegraphics[width=14cm]{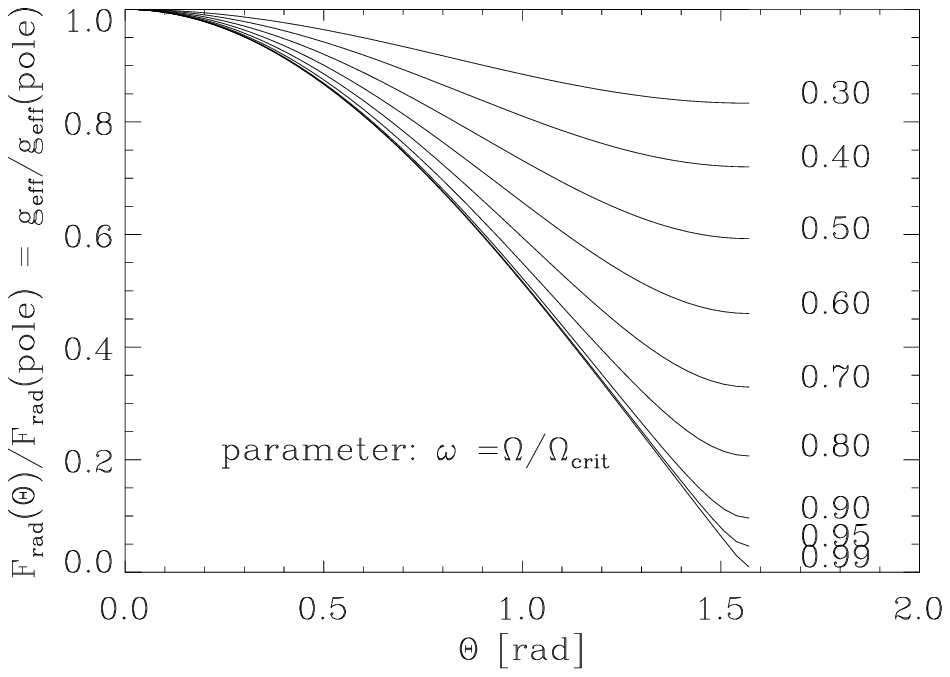}  
 \caption{The radiative flux $F(\Theta)/F(0)$ as a function of 
 the co-latitude $\Theta$ for several values of $\omega$.}
\label{fig:fluxtheta}
\end{figure}

 The radiative flux $F(\Theta)$  compared to the polar flux  is shown in figure \ref{fig:fluxtheta} 
 for several values of $\omega$. 
 These relations are used to calculate the ratio 
  
  \begin{equation} \label{eq:Psi}
  \Psi (\omega) \equiv \Ltot / \Lp  \equiv \Ltot / 4 \pi \Rp^2 \sigma \Tp^4.
  \end{equation}

 The effective gravity is 
 $\geff (0)=G\Meff/\Rp^2$ at the poles and  
 
 \begin{equation}  \label{eq:geffeq}
 \geff(\pi/2) = G\Meff/\Req^2 - \Omega^2 \Req = G\Meff/\Req^2 (1 - \omega^2)
 \end{equation}
 at the equator. This results in a ratio 
 
 \begin{equation} \label{eq:Tratio}
 (\Teq/ \Tp)^4 = (\Rp/\Req)^2 (1-\omega^2) = \frac{(1-\omega^2)}{(1 + 0.5 \omega^2)^2}
 \end{equation}
 where the temperatures are the effective temperatures. From now on we will use \Tp\ and \Teq\ to describe the {\it effective} temperatures at the poles and at the equator.
 
 The dependence of the ratio $(\Teq/\Tp)^4$ on $\omega$ is shown in the middle panel of Figure \ref{fig:equator-pole}.
  The luminosity ratio $\Psi(\omega) = \Ltot / \Lp$  with  $L_{\rm p } \equiv 4 \pi \Rp^2 \sigma \Tp^4$, is shown in the lower panel of Figure \ref{fig:equator-pole}. 
  It reaches a minimum of 0.6 at critical rotation. The function can be described accurately with
  the power law approximation in Eq. \ref{eq:AppPsi} given in Appendix A.
  
  Predicted evolutionary tracks for rotating and non-spherical stars are characterized by luminosity and ``mean temperature'' \Tmean\, which is defined by the surface mean flux (see, for example, \citealt{MeynetMaeder97,Ekstrom12,Georgy13})

\begin{equation} \label{eq:Tmean}
\sigma \Tmean^4 = L / S
\end{equation}
where $S$ is the total surface of the star. The surface of a rotationally distorted star can be expressed as

\begin{equation} \label{eq:s}
S(\Rp,\omega) = 4 \pi \Rp^2 s(\omega)
\end{equation}
where $s(0)=1$ for a non-rotating star and $s(1)=1.21$ for a star rotating at break-up speed.
The function $s(\omega)$ can be expressed in terms of a  power law, which is given by Eq. \ref{eq:sapprox} in the Appendix.

\begin{figure} 
  \includegraphics[width=16cm]{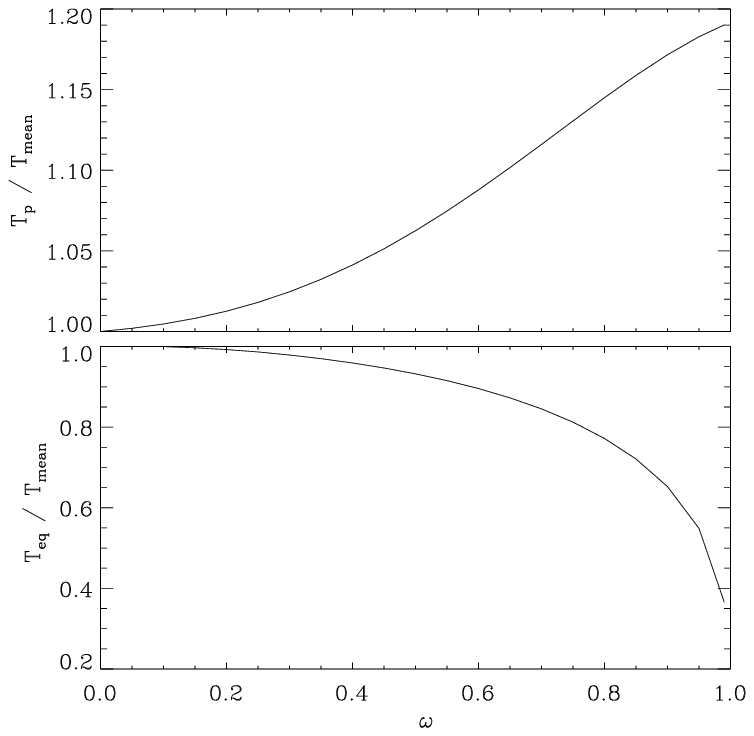}  
 \caption{The ratios $\Tp / \Tmean$ (top) and $\Teq / \Tmean$ (bottom) as a function of $\omega$,
 where $\Tmean$ is the value of the effective temperature for a rotating star, defined by Eq. \ref{eq:Tmean}.
 }
 \label{fig:TeqTmean}
\end{figure}

Combining Eqs. \ref{eq:Tmean} and \ref{eq:s} with Eqs. \ref{eq:Psi} and \ref{eq:Tratio} results in a relation between \Tmean\ and \Teq\ of

\begin{equation} \label{eq:TmeanTeq}
\Tmean^4 = \frac{\Psi(\omega)}{s(\omega)} 
\frac{(1 + 0.5 \omega^2)^2}{(1-\omega^2)} \Teq^4
\end{equation} 
Figure \ref{fig:TeqTmean} 
shows the ratios $\Teq / \Tmean$ and $\Tp/ \Tmean$ as a function of $\omega$. 

\subsection{Triggering the instability in rotating post-main sequence stars}  \label{sec:postMS}

In this section we investigate the triggering of instabilities in rotating
post-main sequence stars of different initial masses with Galactic and SMC metallicities.

The evolution of a star when it leaves the main sequence is described by its initial mass \Mi\ and its properties at the terminal age main sequence (TAMS): \Mtams, \Ltams\ and \Rtams\
and its surface H abundance \Xhtams. 

Values of these parameters at the TAMS for $32 < \Mi / \Msun < 85$ stellar models are given in Table 2, based on the Geneva evolutionary models for rotating stars that include mass loss \citep{Ekstrom12,Georgy13}. The adopted initial relative angular velocity in these models is $\omega=0.4$.  Table 2 lists mass, luminosity, mean surface temperature, radius, surface H-abundance, angular velocity, relative angular velocity $\omega$, and the rotational velocity at the equator, at the point when the star leaves the main sequence and starts its redward evolutionary across the HRD.

 The data in Table 2 show that $\omega$ decreased drastically during the main sequence for the stars with galactic
 metallicity, $Z=0.014$,  as well as the $85 \Msun$ star with low metallicity, $Z=0.002$.
 This is due to the high mass loss in these stars, which has carried off a large fraction of their angular momentum. We will use the luminosity and mass of these models at the TAMS as a starting point for our predictions of the triggering of the S Dor instability.

  \begin{deluxetable*}{l c c c c c c c c c c c }
 \tablewidth{0pt}
 \tabletypesize{\footnotesize}
 \tablecolumns{11}
 \tablecaption{Properties of stars at the TAMS}
 \tablehead{
     \colhead{$Z$} & \colhead{$\omega_i$} &  \colhead{$M_i$} & \colhead{$M$} &
     \colhead{log($L$)} & \colhead{log($L$/$M$)} & \colhead{log($T_{\rm mean})$} & \colhead{$R$} & \colhead{$\Xh$} &
     \colhead{log($\Omega$)} & \colhead{$\omega$} & \colhead{$v_{\rm rot}$}  \\
     \colhead{} & \colhead{} & \colhead{$\Msun$} & \colhead{\Msun} & \colhead{\Lsun} & \colhead{\Lsun/\Msun} & \colhead{K} & \colhead{\Rsun} &  
     \colhead{ } & \colhead{rad s$^{-1}$} & \colhead{} & \colhead{km s$^{-1}$}
} 
    \startdata
    0.002  & 0.4 &  32 &  31.2 & 5.561 &4.067 & 4.503 &  19.8 & 0.69 & -4.733 &  0.465 & 254  \\
    0.002  & 0.4 &  40 &  38.7 & 5.747 &4.159 & 4.508 &  19.3 & 0.68 & -4.939 &  0.348 & 190  \\
    0.002  & 0.4 &  60 &  56.4 & 6.078 &4.327 & 4.552 &  28.8 & 0.60 & -5.188 &  0.213 & 130  \\
    0.002  & 0.4 &  85 &  77.2 & 6.335 &4.447 & 4.642 &  25.5 & 0.46 & -5.608 &  0.006 &  44  \\
    0.014  & 0.4 &  32 &  28.2 & 5.595 &4.145 & 4.450 &  26.3 & 0.55 & -5.942 &  0.046 &  21  \\
    0.014  & 0.4 &  40 &  32.0 & 5.779 &4.274 & 4.479 &  28.5 & 0.43 & -5.821 &  0.065 &  30  \\ 
    0.014  & 0.4 &  60 &  38.5 & 6.047 &4.462 & 4.713 &  13.2 & 0.16 & -5.949 &  0.014 &  10  
    \enddata
    \label{tbl:TAMS}
 \end{deluxetable*}

As stars expand during post-MS evolution, the luminosity hardly changes. We can therefore assume that
$L(t) \simeq \Ltams$. This is confirmed by the stellar evolution models.

If the star rotates when it leaves the TAMS the surface rotation rate $\Omega(t)$ 
will decrease as the radius increases. If the envelope is rotationally decoupled from the 
core and the expanding inflated layers, because of the large density contrast and the absence of convection outside 
the core, the angular momentum of the envelope is conserved. If stellar winds during the fast expansion in the early
post-MS phase do not significantly reduce the mass of the envelope, $\Omega$ decreases as  $R^{-2}$.

\begin{figure} 
  \center
  \includegraphics[width=14cm]{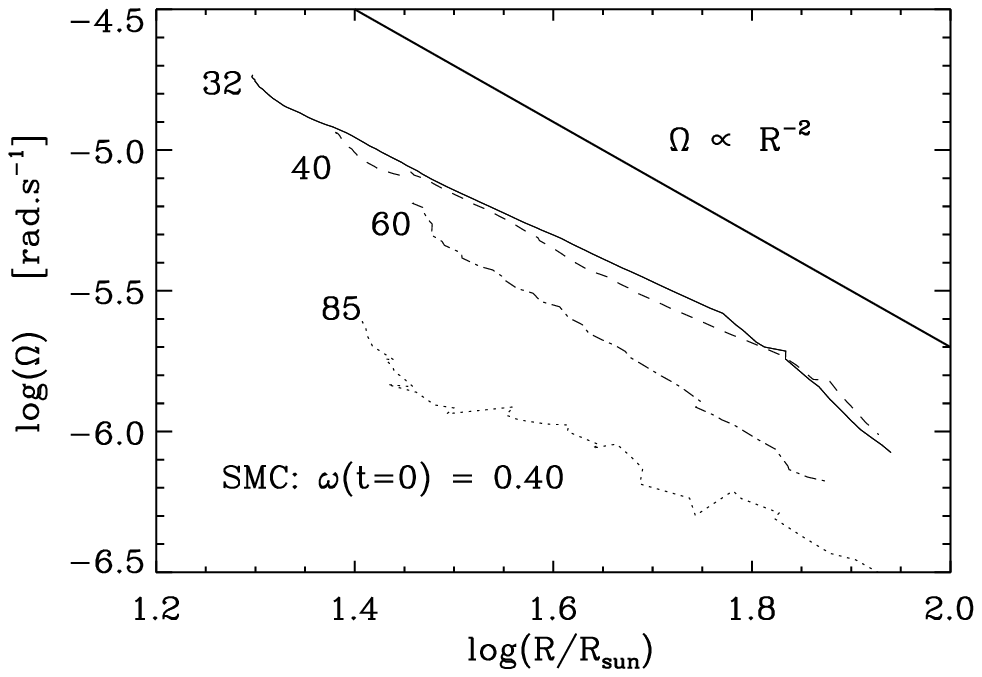}  
 \caption{The variation of $\Omega$ as function of $R$ during the post-MS expansion of rotating stars 
 of $M_{\rm i}=32,$ 40, 60 and 85 $\Msun$ with metallicity $Z=0.002$ predicted by \citet{Georgy13}.
 The models have an initial rotation rate of $\omega = \Omega/\Omegacrit = 0.4$. The straight line in the upper right hand corner shows the predicted slope of $\Omega_{\rm pred}\simeq R^{-2}$.  
 }
 \label{fig:omega-geneva}
\end{figure}

This predicted decrease of $\Omega \propto R^{-2}$ is confirmed by the evolution models of rotating massive stars. 
Figure \ref{fig:omega-geneva} 
 shows the variation of $\Omega$ as a function of $R$ during the post-MS expansion
of several massive stars with $Z=0.002$ \citep{Georgy13}. The initial rotation rate is $\omega = 0.4$, 
which implies $\vrot \approx 300-400$ km s$^{-1}$ at the ZAMS for stars of $32 < M/\Msun < 85$. The figure shows that
$\Omega$ evolves approximately as $R^{-2}$ when the star expands.

We have argued above that the structure of the star is determined by $\omega = \Omega/\Omegacrit$,
with $\Omega \propto R^{-2}$ and $\Omegacrit \propto R^{-3/2}$ for constant mass. 
As a result, during the fast post-MS expansion at nearly constant mass $\omega$ varies as 

\begin{equation}
\label{eq:omegavar}
\omega \propto R^{-1/2} \propto \Teff.
\end{equation}
These proportionalities  are only valid if $\omega < 0.4$, i.e. when the rotational distortion of the star is small, 
with $\Req/\Rp< 1.08$. We see that $\omega$  decreases as the star expands and $\Teff$ decreases.

The S Dor instability is triggered at the {\it equator} of a rotating star when 
\begin{equation} \label{eq:criteq1}
 \frac{\kapmax(\Teq) \sigma \Teq^4}{c}=0.9 \frac{GM(1-\omega^2)}{\Req^2}
\end{equation}

The von Zeipel theorem states that  
$\Teq^4 /\Tp^4 = g_{\rm eq}/g_{\rm p}$ with $g_{\rm eq}=GM(1-\omega^2)/\Req^2$ and $g_{\rm p}=GM/\Rp^2$. Substitution of this into Eq. \ref{eq:criteq1} together with

\begin{equation} \label{eq:Lp2}
\sigma \Tp^4 \Rp^2 \equiv  \frac{\Lp}{4 \pi} =  \frac{L}{4 \pi \Psi(\omega)}
\end{equation}
(see Eq. \ref{eq:Psi}) results in a triggering criterion for the equator of 
$(L/M) \ge (L/M)_{\rm crit}$, with

 \begin{equation} \label{eq:LMcriteq}
    \left( \frac{L/\Lsun}{M/\Msun}\right)_{\rm crit, eq} =
     \frac{0.9 \times 4 \pi G c \Psi(\omega)}{\kapmax(\Teq)} \cdot \frac{\Msun}{\Lsun} =
     \frac{1.17 \times 10^4 \Psi(\omega)}{\kapmax(\Teq)}.
  \end{equation}
This is equivalent to the criterion for non-rotating stars (see eq. \ref{eq:LMcrit}).
 
A similar analysis for the pole shows that the instability at the polar regions will be triggered if

 \begin{equation} \label{eq:LMcritp}
    \left( \frac{L/\Lsun}{M/\Msun}\right)_{\rm crit, p} = 
     \frac{0.9 \times 4 \pi G c \Psi(\omega)}{\kapmax(\Tp)} \cdot \frac{\Msun}{\Lsun} =
     \frac{1.17 \times 10^4 \Psi(\omega)}{\kapmax(\Tp)}.
  \end{equation}
The definitions for these two criteria are identical, apart from the fact that the absorption coefficients are at different temperatures.  Using the definition of
$k=\kapmax/\kapmaxUF$ and $\kapmaxUF=0.313/f$ (Eq.\ 6), these equations can also be expressed as

\begin{equation} \label{eq:LMcritrot}
   \left( \frac{L/\Lsun}{M/\Msun}\right)_{\rm crit} = 
     \frac{3.73 \times 10^4 \Psi(\omega) \times f(T)}{k(Z, T, \Xh)}.
\end{equation} 
for $T=\Teq$ and $T=\Tp$ respectively, with $0.75<k< 1$  (Eq. \ref{eq:k}).

\begin{figure} 
  \centering
  \includegraphics[width=14cm]{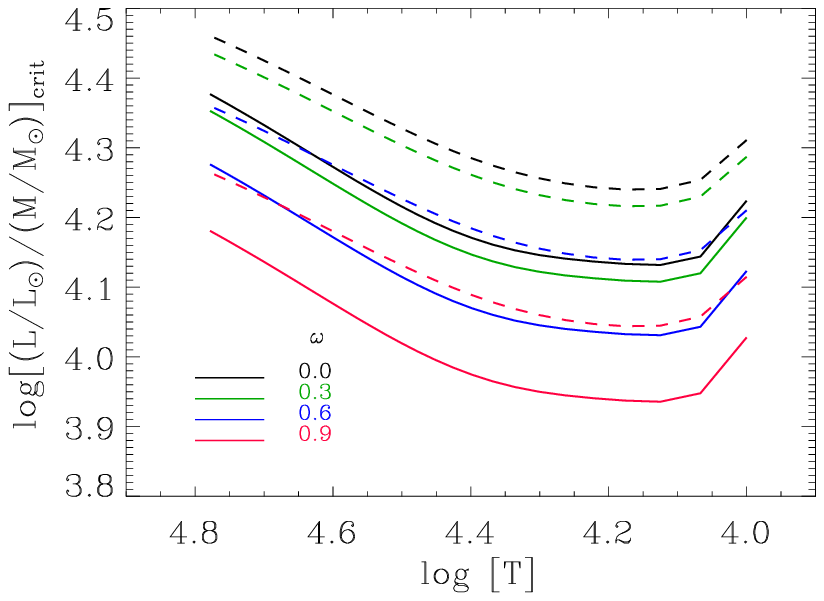}  
 \caption{The dependence  of $\LMcrit$ on $T$ for rotating models of $\Xh=0.74$ and values of $\omega$=0 (black), 0.3 (green), 0.6 (blue) and 0.9 (red). 
 Full lines are for $Z=0.02$ and dashed lines for $Z=0.002$.
 For $\Xh < 0.74$ the values of log$(L/M)_{\rm crit}$ have to be corrected by a factor log$(1/k) \le 0.125$ (Eq. \ref{eq:k})
 }
 \label{fig:LMcrit}
\end{figure}

The values of  $(L/M)_{\rm crit}$ are plotted in Figure \ref{fig:LMcrit}
for $\Xh=0.74$ and $\omega=0$, 0.3, 0.6 and 0.9. For other values of \Xh\ these would need to be increased by   a factor $1/\kztx \simeq 1$  (Eq. \ref{eq:k} and Fig. 4).
At high temperatures they decrease with decreasing \Teff, reach a minimum at $T \simeq 12500$, and then increase steeply
towards lower temperatures.

 \subsection{Implications}  \label{sec:implications} 
 
 The criteria described above imply that:

1. If a non-rotating star expands in the post-MS phase with nearly constant $L/M$, \Teff\ decreases and $f(T) = g_E/g_N$  increases as long as $T > 12500$ K, and hence the instability will be triggered when $\LMcrit$ reaches $L/M$.

2. The minimum $L/M$ ratio required to trigger the instability in non-rotating stars is given by Eq.
\ref{eq:LMcrit} at T=12500 K, which results in 

\begin{equation}
   \left( \frac{L / \Lsun}{M / \Msun}\right)_{\rm min} \ge 1.4 \times 10^4/ \kztx
\end{equation}
for $Z=0.02$ stars and $1.8 \times 10^4/ \kztx$
for $Z=0.002$ stars, with $k$  defined by Eq. \ref{eq:k} and shown in Fig. 4. 
The evolutionary models of \cite{Ekstrom12} and \cite{Georgy13} suggest that this happens for a minimum initial mass of $\Mi \simeq 60\Msun$ for stars with $Z=0.014$ and 
$\Mi \simeq 85\Msun$ for stars with $Z=0.002$. Given that stars with much lower presumed initial masses have been observed to show S Dor variations, this implies that rotation {\it must} play a critical role in triggering the S Dor instability.

3. In a rotating star the equatorial temperature is lower than the polar temperature. As \Tmean\ decreases during the post-MS expansion with $\Teq < \Tmean$  while $\Tp > \Tmean$, the equator will reach the instability criterion while the polar regions are still stable. This will produce a higher mass-loss rate  from the equatorial regions than from the polar regions. However, the ejection velocity, which is expected to scale with the local escape velocity, $v_{\rm esc}(\Theta)$, will be
higher at the poles. These two effects both play a role in shaping LBV nebulae (see Section 6).
 
 
\subsection{Quantitative predictions} \label{sec:predictions}

In the previous section we derived the condition for the triggering of the instability at the equator of a rotating or non-rotating star (see Eqs.\ \ref{eq:LMcriteq}, \ref{eq:LMcritp}, \ref{eq:LMcritrot}). We also showed that the instability will always be triggered first at the equator (unless the star moves from right to left in the HRD). We can now use these expressions, together with the ratio $\Teq/\Tmean$ (Eq. \ref{eq:TmeanTeq}),
to calculate the location of stars in the HRD when
the instability is triggered at the equator. 
The result is shown in Figure 10 for $\Xh=0.74$ (for $\Xh<0.74$ these must be corrected by $0 < \log(1/\kztx) \le +0.125$).

\begin{figure}[ht!] 
\includegraphics[width=17cm]{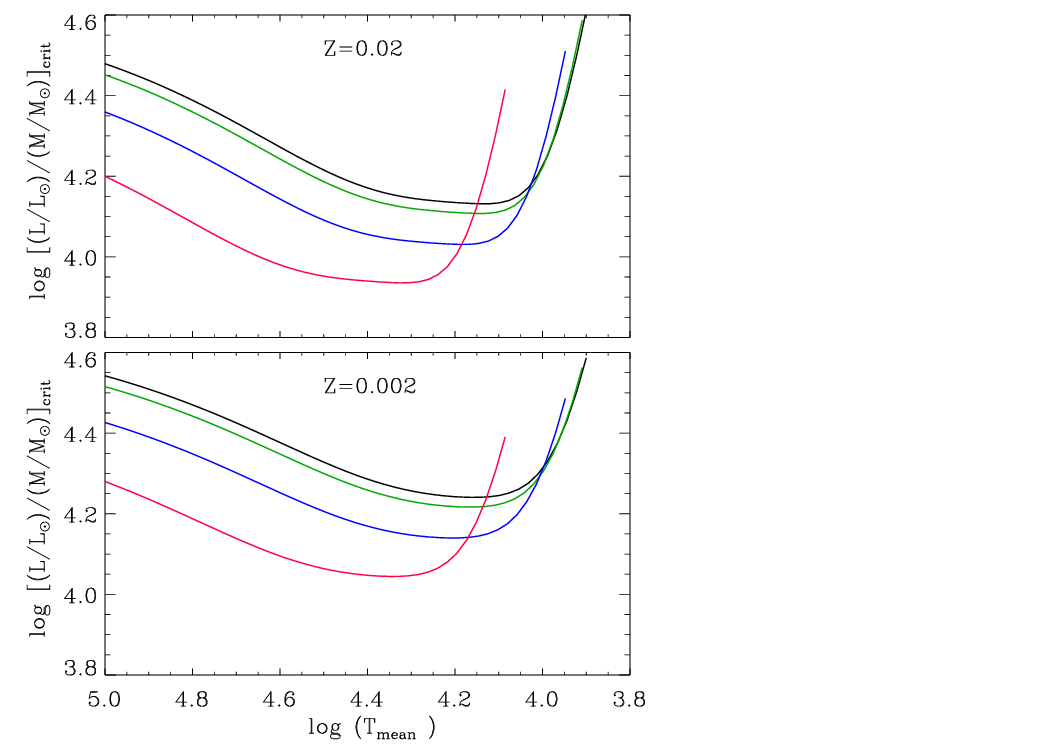}
\caption{The predicted critical values of $L/M$, where the equator of a rotating star becomes unstable as a function of the mean temperature $\Tmean = \Teff$,  for $\Xh=0.74$ and 
$Z=0.02$ (upper panel) and 0.002 (lower panel) and for different rotation rates $\omega=0$  (black), 0.3 (green), 0.6 (blue) and 0.9 (red).  For $\Xh < 0.74$ the data have to be corrected by $0 < \log(1/\kztx) < 0.125$.
}
\label{fig:LMTmean}
\end{figure}

For a given value of $L/M$  the value of \Tmean\ where instability is triggered increases with increasing rotation rate. In other words, if a star expands after the main sequence phase the instability occurs earlier in a rapidly rotating star than in a slowly rotating star.

 Figure \ref{fig:LMTmean}
 shows that at a given  $\Tmean$ and $\omega$ the high-metallicity stars become unstable at a lower value of $L/M$ than low-metallicity stars. This is because the absorption coefficient $\kapatm$ and the 
 resulting radiation pressure is higher in the atmospheres of high-metallicity stars. Alternatively, at a given value of $L/M$ the high-metallicity post-MS stars reach the stability limit at a higher $T_{\rm eff}$ (i.e. earlier in their evolution) than low-metallicity stars.
 

\section{Models of rotating LBVs in the instability strip}

\subsection{Physical Properties of LBVs}

To numerically quantify the potential effects of rotation on the triggering of the instability of LBVs in their post-MS phase we need approximate values of $L/M$ and $\Xh$. We begin by adopting theoretical stellar parameters that best match the physical state of stars in the S Dor instability strip. To do this we use the Geneva stellar evolutionary models which include the effects of mass loss, computed for both non-rotating and rotating stars at solar ($Z=0.014$) and SMC ($Z=0.002$) metallicities \citep{Ekstrom12,Georgy13}, comparing these evolutionary tracks to the location of the S Dor instability strip (the observed diagonal distribution of LBVs in their hot state; see Figure \ref{fig:fig1}) on the HRD. In this manner we are able to use the Geneva models as a guide to dictate what the parameters of a model LBV should be when it arrives at the instability strip.

\begin{figure*}[ht!]
\center
  \includegraphics[width=.49\textwidth] {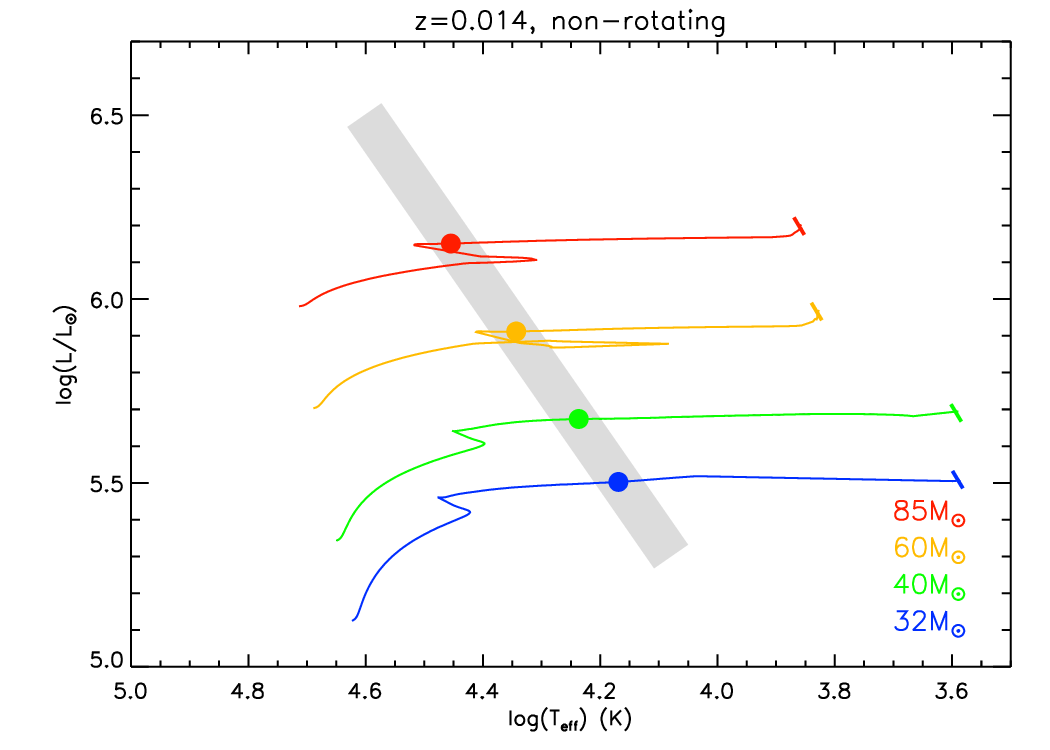}  \includegraphics[width=.49\textwidth] {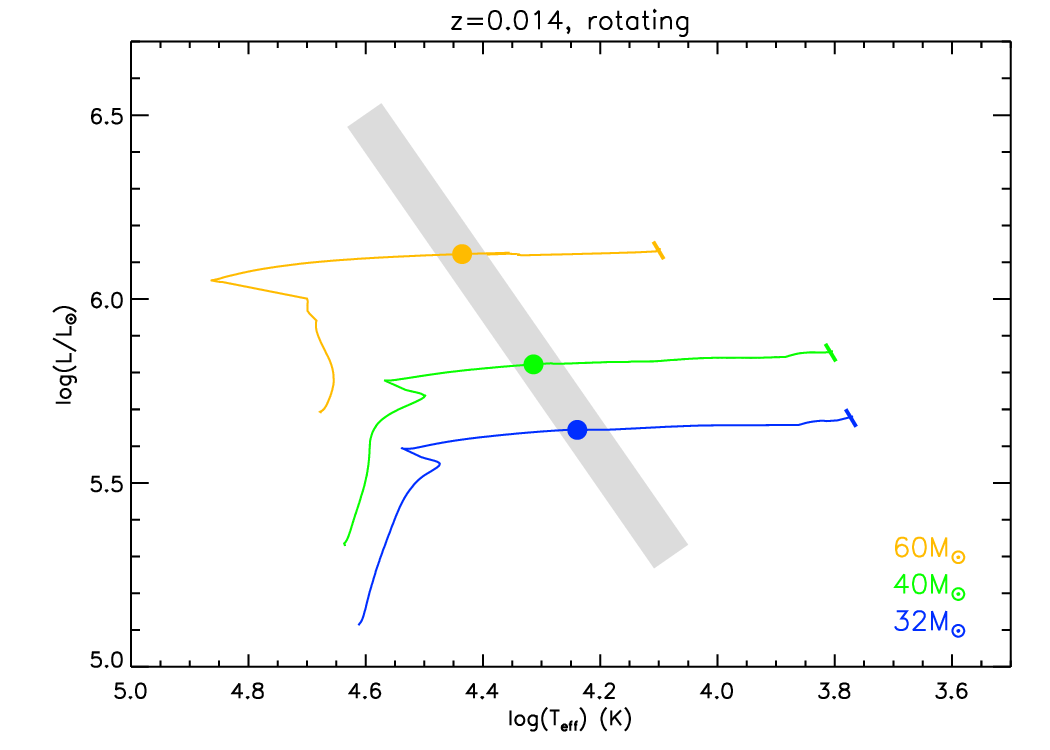}
  \includegraphics[width=.49\textwidth] {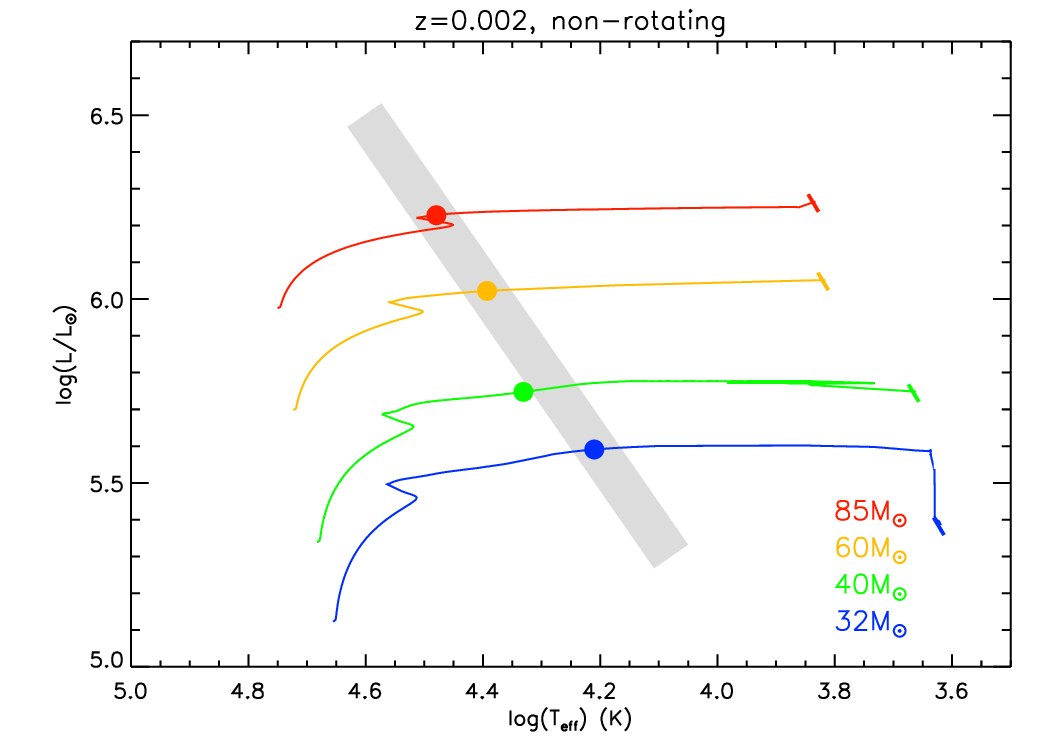}
  \includegraphics[width=.49\textwidth] {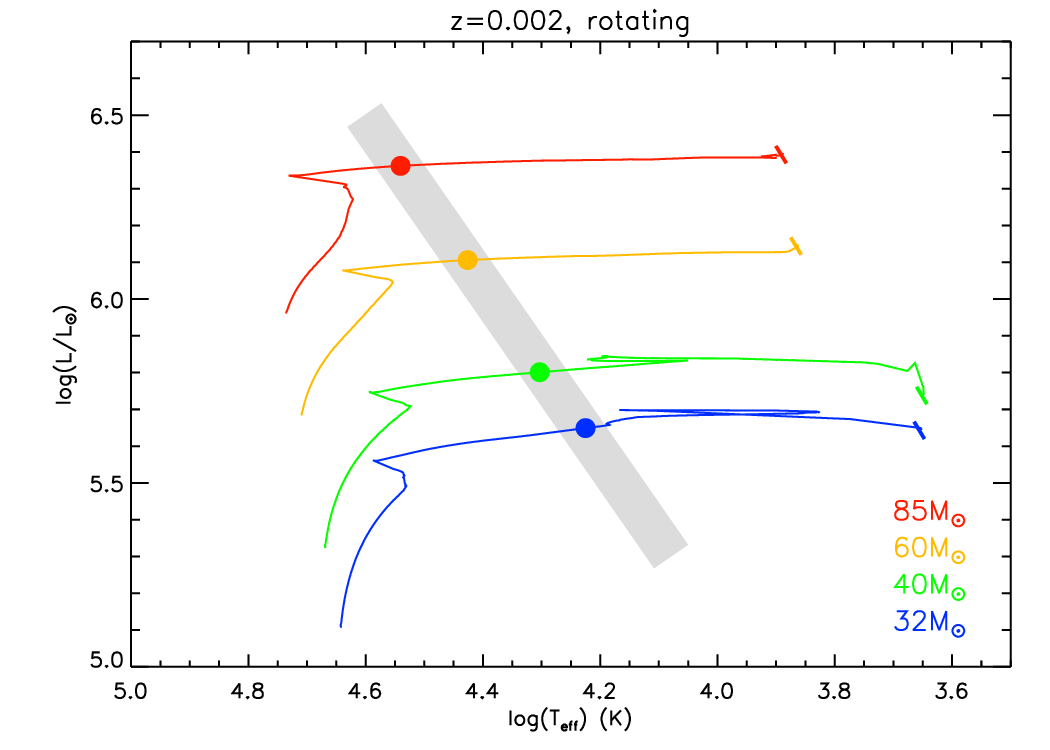}
 \caption{Locations of our model LBVs on the HRD, as compared to the non-rotating (left) and rotating (right) Geneva evolutionary models for solar (top) and SMC (bottom) metallicity. The location of the S Dor instability strip is overplotted in grey; the parameters for our model LBVs were chosen based on when the evolutionary tracks first cross the instability strip during their post-MS evolution. Hash marks at the end of the evolutionary tracks indicate that these tracks have been terminated at their coldest point to focus on the stars' early post-MS evolution and avoid confusion with the later evolutionary stages of these stars, which in some cases cross leftward across the HRD and enter a Wolf-Rayet phase. The evolutionary track for a M$_i=85$M$_{\odot}$ solar-metallicity rotating star never crosses the S Dor instability strip; as a result, we do not include a M$_i=85$M$_{\odot}$ solar-metallicity rotating model LBV in this work.}
 \label{fig:evoltracks}
\end{figure*}

The results of this comparison are illustrated in Figure \ref{fig:evoltracks}. For completeness we estimated our model LBVs' physical properties using both rotating and non-rotating Geneva evolutionary models, in order to account for the potential effects of rotation on the stars' pre-LBV evolution. This enables us to use a wider range of physical properties (in particular the $L/M$ ratio, which is impacted by rotational mixing and mass loss effects on the main sequence) for stars that have arrived at the S Dor instability strip. As noted previously, the rotating Geneva models adopt $\omega=\vrot / \vcrit = 0.4$ at the ZAMS for each star.

Taking the 32, 40, 60 and 85\Msun\ evolutionary tracks, we note the first post-main-sequence step in each stellar model's evolution that corresponds with the position of the S Dor instability strip and adopt the physical properties of the stars at this evolutionary phase. These parameters are: the luminosity $L$, the current mass $M$, the current surface 
H abundance \Xh, and the mean surface temperature \Tmean, where $\sigma \Tmean^4$ is the surface averaged mean radiative flux (eq. \ref{eq:Tmean}). While these parameters are specifically chosen to represent stars that have reached the post-main-sequence S Dor instability strip, it is interesting to note that the values of $L$, $M$, and $X_H$ are practically identical to the state of these stars at the TAMS (see Table 2); the only notable difference is in the stars' cooler \Tmean\ at the instability strip.

 As noted in Figure \ref{fig:evoltracks}, the rotating solar-metallicity Geneva models predict an evolution for the 85 \Msun\ star that never reaches the S Dor instability strip.  This is a consequence of the high mass-loss rate in this model caused by fast rotation during the main sequence phase, which leads to an almost immediate leftward evolution of the star on the HRD \citep{Ekstrom12}. Therefore we do not include a $M_i=85M_{\odot}$ solar-metallicity rotating LBV model in our analyses. 

The parameters adopted based on the Geneva models are included in Table 3. Adopting these as a starting point, we can then estimate the effects of a variety of rotation rates on these model LBVs' physical properties, the potential
triggering of the instability and, if the instabiltiy is triggered,  its influence on the star's mass loss, circumstellar environment, and potential observable properties.

 \subsection{Numerical methods}

 Using the above parameters from the Geneva models together with the preceding theory for defining the structure of a rotating star and the triggering criteria for the instability, we can now numerically determine whether a broad range of model stars on the S Dor instability strip produce polar or equatorial instabilities and how this might manifest itself observationally\footnote{For the Galactic stars we ignore the small difference in metallicity between $Z=0.014$ adopted for  the Geneva models and $Z=0.02$ used by UF98 to derive the values of \kapmaxUF\ and \fT.}.

To begin, we apply several different values for the star's present rotation rate ($\omega$  = 0.0, 0.3, 0.6 and 0.9) by using Eq. 12 to set the corresponding \Req/\Rp\ ratios (\Req/\Rp = 1.000, 1.045, 1.18, and 1.405 respectively).  By adopting $L$ and \Tmean\ from the evolutionary models, we implicitly assume that the total surface of the star at that evolution phase is independent of its rotationally-distorted shape. Based on this we can use Eqs. 18 and 19 to define the surface of the star and determine a value for \Rp, which in turn gives us \Req\ for a specific value of $\omega$ (We note that the initially rotating SMC model of $\Mi=85$ \Msun\ has such a high $L/M$ ratio that it reaches $\Gamp>1.0$ if $\omega=0.9$. This model is unphysical and is therefore omitted from further analysis).

To determine the corresponding polar and equatorial temperatures, \Tp\  and \Teq, we first use our power law approximation for $\Psi(\omega) \equiv \Ltot / \Lp$ (Eq. A2 in the appendix) and $L=\Ltot$ from our model star to determine \Lp. Combining that with \Rp\ we can determine \Tp, and use Eq. 17's relationship between \Teq/\Tp\ and $\omega$ to determine \Teq.

With these values of $T_p$ and $T_{eq}$ we can now determine if the physical properties of our model star satisfy the instability criteria. We must first determine \kapmax\ at the pole and equator, which can be done using our Eq. 6 and 7 and our power law fits $f(T_{\rm eff})$ = \gE/\gn\ from UF98 for both $Z=0.02$ and $Z=0.002$ (given in Eq. A1 of the appendix). With \kapmax\ defined at both the pole and equator of our star, we can then use the criteria given in Eqs. 24 and 25 to determine whether our model star will be unstable  at the equator or the pole or both.

We want to estimate the velocity of the resulting wind from the poles and the equator. 
If we assume that the wind is accelerated outwards by line-driving we can use the theory
and observations of radiation driven winds from hot stars, which both show that $\vinf \propto \vesc$ (e.g. \citealt{LamersSnowLindholm95, Lamers99}).
The escape velocities at the equator and poles are  \begin{equation} \label{eq:vesc}
 \begin{split}
  \vescp = \sqrt{2 G \Meffp / \Rp} \\
  \vesceq = \sqrt{2 G \Meffeq (1-\omega^2)/ \Req} 
 \end{split}
 \end{equation}
 with $\Meff=M(1-\Gammae)$ and Eddington factors at the equator and the poles, \Gameq\ and \Gamp, 
 defined by 
 
 \begin{equation} \label{eq:gamma}
 \begin{split}
     \Gameq = \sigmae \sigma \Teq^4 \Req^2/ c G M \\
     \Gamp = \sigmae \sigma \Tp^4 \Rp^2/ c G M
\end{split}
\end{equation}
 with $\sigmae = 0.200(1+\Xh)$ (Notice that this results in different values of $M_{\rm eff}$ at the equator and poles due to \Gammae's dependence on \Teff\ and $R$).
 We adopt the empirical ratios 
 $v_{\infty}/\vesc \simeq 2.6$ if $T \ge 21 000$K,
 1.3 if $10000 \le T < 21000$ K, and 0.7 if $T<10000$K  \citep{LamersSnowLindholm95}. The bistability jumps at 21000 and 10000 K are caused by changes in the ionization states of the dominant driving elements \citep{Vink99}. We should note that for instability-driven winds $v_{\infty}$ may be smaller than for radiation-driven winds, but \vinf\ is still expected to be proportional to $\vesc$.

 
 \subsection{Models for rotating LBVs}
 
 The resulting models for rotating LBVs are listed in Table 3. 
 The following parameters are listed: the initial mass $\Mi$, the present mass $M$ (the mass of the star at the evolutionary step we adopt for our models), the luminosity $L$, the key ratio $L/M$,
 the mean surface temperature $\Tmean$, the hydrogen surface abundance $\Xh$, the adopted rotation rate $\omega$,
 the polar and equatorial radii $\Rp$ and $\Req$, the polar and equatorial 
 effective temperatures $\Tp$ and $\Teq$, and the equatorial rotation speed $\vroteq$. We also list the values of 
 \vescp\ and \vesceq. The column ``Instab?"  indicates 
 whether the instability is triggered at the equator or poles, according to the criteria of Eq. \ref{eq:LMcriteq} and \ref{eq:LMcritp}.
 This is indicated in the table by {\it ``N"} (not triggered) or {\it ``Y"} (triggered). $Y_{\rm eq}$ and $Y_{\rm p}$ indicate that the triggering occurred {\it only} at the equator or pole respectively.
 
\begin{figure} [ht!]
  \center
\includegraphics[width=0.49\textwidth]{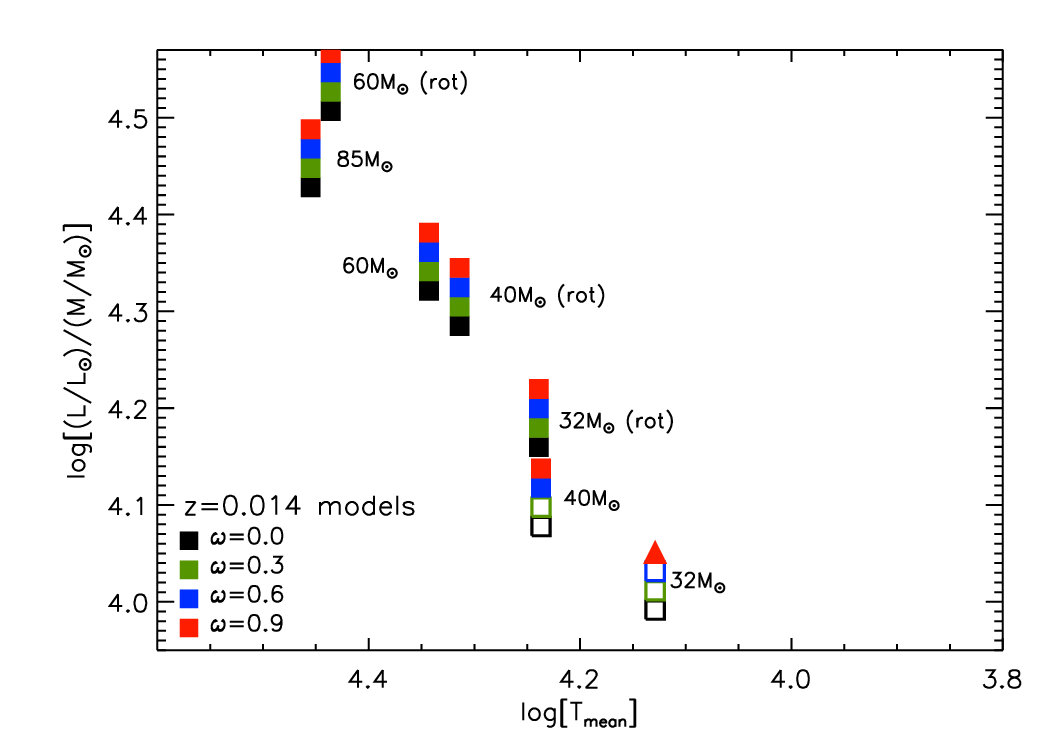} 
 \includegraphics[width=0.49\textwidth]{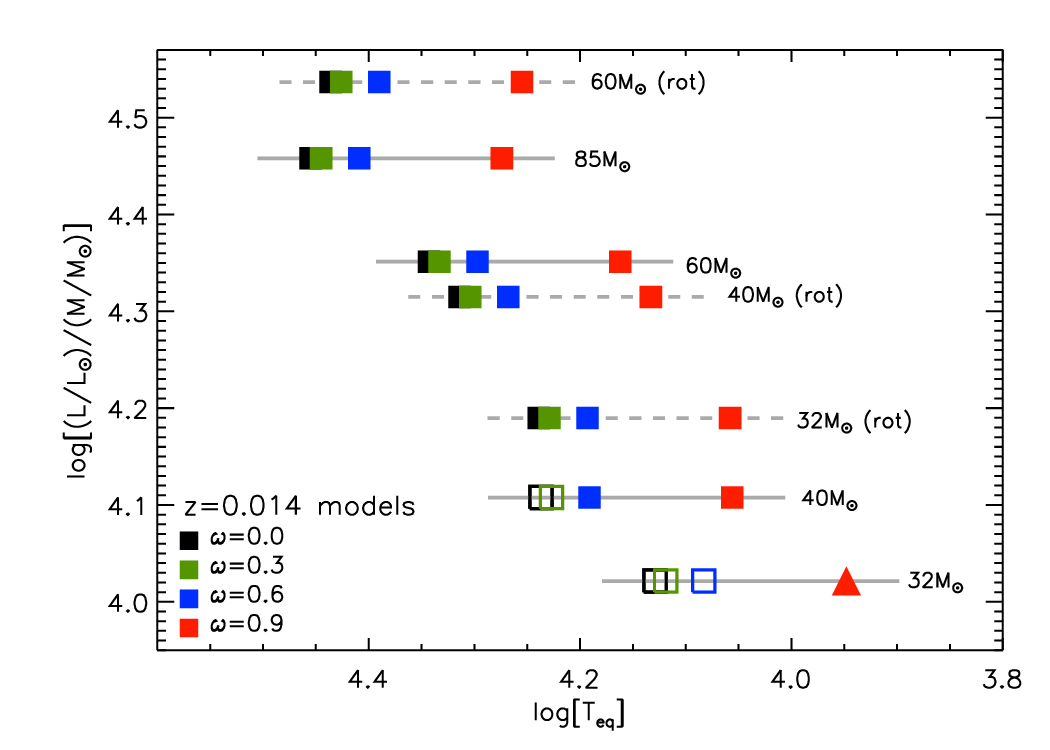} 
 \caption{The effect of rotation on stability for stellar models with Galactic metallicity (listed in Table 3). {\it Left}: $\Tmean$ versus the parameter $L/M$ for various values of $\omega$ (indicated by color according to the legend). Filled squares indicate that both the poles and equator of the star have passed the stability limit and are unstable. Open squares indicate that the model is stable at both the poles and the equator. Filled upward triangles indicate models that are unstable only at the poles (seen only in the Milky Way models), and filled downward triangles indicate models unstable only at the equator (seen only in the SMC models). Initial masses of each model set are labeled. {\it Right}: $\Teq$ versus $L/M$ of the same models, illustrating the cooler equatorial temperatures produced at higher values of $\omega$. Symbols are the same as in the top panel, but Geneva models with initial rotation rates $\omi=0$ are connected by solid gray horizontal lines while models with $\omi=0.4$ are connected by dashed gray lines; models on the same connecting line have the same initial mass, as labeled.}
\label{fig:Gal}
\end{figure}

\begin{figure} [ht!]
  \center
  \includegraphics[width=0.49\textwidth]{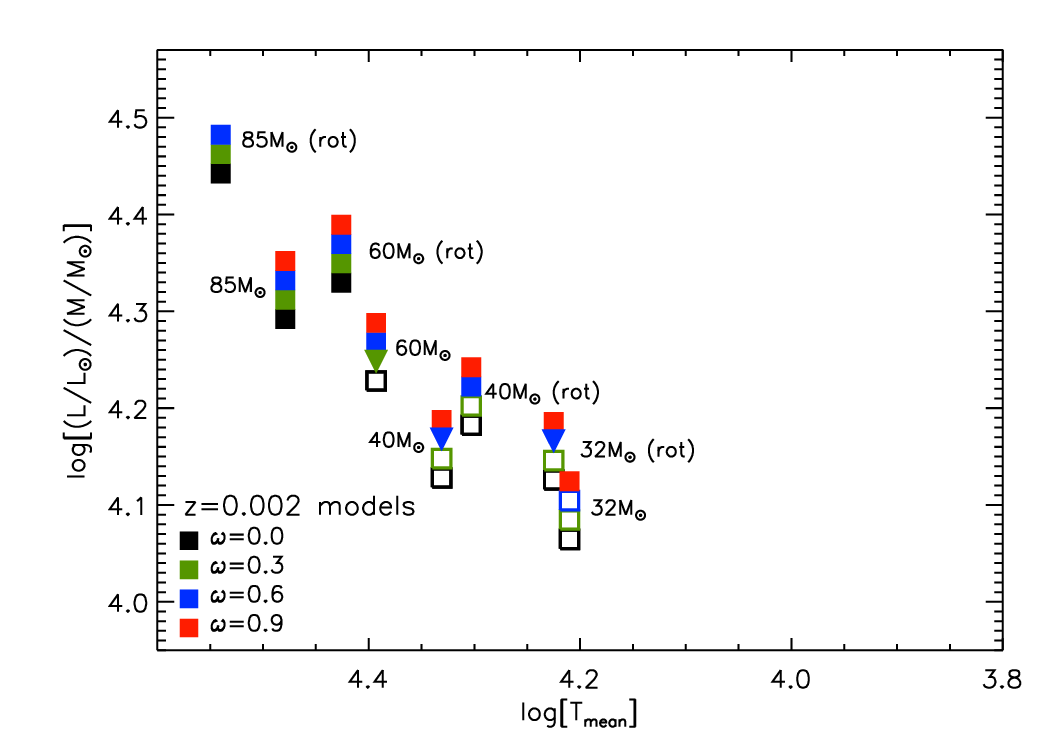} 
  \includegraphics[width=0.49\textwidth]{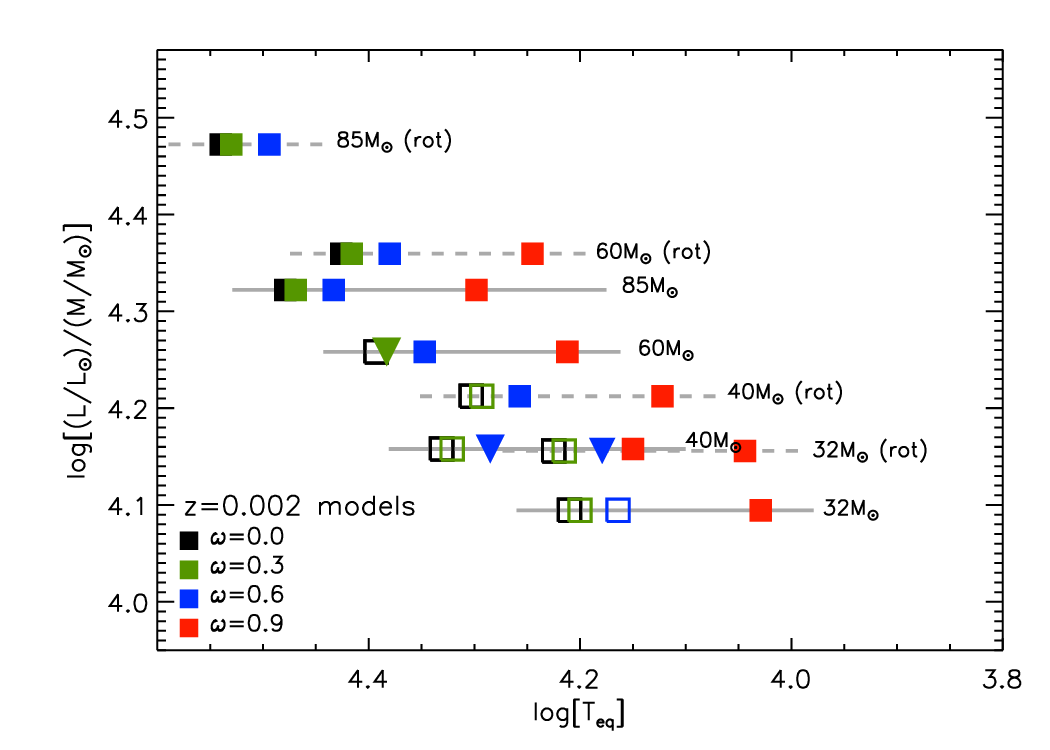}  
 \caption{ As in Figure \ref{fig:Gal} for models with SMC metallicity.
  }
\label{fig:SMC}
\end{figure}

 Figures \ref{fig:Gal} and \ref{fig:SMC} show the results of the evolution models listed in Table 3 in terms of $\Teq$ versus the parameter $(L/M)$ for several values of $\omega$. The relation between $\Teq$ and this parameter was derived in \S 5.2.
 

 The data in Table 3 and in Figs. 
 \ref{fig:Gal} and \ref{fig:SMC} show some general trends:\\
 \begin{enumerate}
  \item Any star that is close to instability at small $\omega$ will be unstable at some higher rotation rate. This instability will first occur at the equator because $\kappa_{eq} > \kappa_p$.
 \item Initially rotating stars have a higher value of $L/M$ when they reach the S Dor strip than non-rotating stars. This is due to their higher $L$ (due to a larger core) and lower $M$ (due to higher mass-loss rates). Similarly, initially rotating stars with Galactic metallicity also have a higher value of $L/M$ when they reach the S Dor strip than their SMC-metallicity counterparts, a result of their higher mass-loss rate and subsequent lower current mass.
 \item Initially non-rotating stars of $\Mi=32$ and 40 \Msun\ with Galactic metallicity have a lower value of $L/M$ when they reach the S Dor strip than those of SMC metallicity; the $\Mi=32~\Msun$ stars do not reach the stability limit unless their rotation rate is as high as $\omega=0.9$.
\item  
     Models with $\Mi=85 \Msun$, regardless of their metallicity or initial rotation treatment, are unstable at all rotational rates. This is due to their high value of  $L/M$. The same is true for models of $\Mi=60 \Msun$, except for the initially non-rotating models with 
     $Z=0.002$ which have a relatively low $\log(L/M)=4.258$. This is due to the star's higher present mass (a consequence of low mass-loss rates), resulting in a lower value 
     of $L/M$ than $60 \Msun$ stars modeled with initial rotation or higher metallicity.
\item  Models of stars with $\Mi=40 \Msun$ are stable if $\omega \le 0.3$ but unstable if $\omega \ge 0.6$. The only exception is the $Z=0.014$ initially rotating model.
\item  If a rotating star passes the stability limit at the equator for some value of $\omega$, the polar region
     may still be stable. This will occur when $\Teq$ is significantly
     lower than $\Tp$ (i.e. at large $\omega$). In our set of models this only occurs for a few low-metallicity models: $\Mi=40$ and 60\Msun\ with $L/M$ derived from non-rotating models, and 32\Msun\ with $L/M$ derived from rotating models.
\item  Only one of our models has a stable equator and an unstable polar region. This is the initially non-rotating
     $Z=0.02$ model with $\Mi=32 \Msun$ and $\log (L/\Lsun)=5.50$ when $\omega=0.9$. 
     The equatorial temperature of this model is so 
     low ($T_{eq}=8800$ K) that $\kapmax$ has dropped to a low value of $<0.5$ cm$^2$/g (see Figure \ref{fig:fT}), 
      resulting in a small radiation pressure. At the same time $\Tp = 13 500$K, which is near the peak of  the 
     \kapmax - curve at 0.85 cm$^{2}$/g.  It is unlikely that this situation will occur during the 
     post-MS evolution of a star, because  the equator will have previously reached the instability limit at a \Teff\ higher than the value of $\Tmean = 13500$K adopted here. 

 \item In rapidly rotating models with high $L/M$ ratios the equatorial wind speed may be higher than that the poles. This occurs when the reduction of \geff\ at the poles by the factor ($1-\Gammae$) is stronger than the reduction at the equator due to rotation. This is the case in some models with $\Mi = 60$ or 85 \Msun\ and $\omega \ge 0.6$. \\

\end{enumerate}    

We realize that the stellar parameters of the models in Table 3 were chosen in such a way that they are, by definition, at or very close to the S Dor strip. Therefore it is not surprising to find that most models are unstable at their equator even at small values of $\omega$. For parameters other than those adopted here, the instability at the equator and poles can easily be calculated from the equations and numerical methods described above and in section \ref{sec:postMS}, or from the curves in Figure 10 with a small correction of $0< \log(k)<0.125$.


\section{Comparison with Observations}
\label{sec:results} 

Our models predict the triggering of S Dor variations by the instability of the photosphere due to radiation pressure at $\tau \le 10^{3}$. Although the mass of the photosphere is small, the  triggered ejection by the subphotospheric layers can be large due to the proximity to the MEL. Observations suggest that the mass-loss rate during the
S Dor variability is of order $\sim10^{-4}$ to $10^{-5}$ \Msunyr\ \citep{Lamers89, HumphreysDvidson94,Dekoter96,Leitherer97,SmithVinkdeKoter04,Vink12}
and the S Dor variations last years to decades. The observed LBV nebulae have a much higher mass, typically about 1 to 10 \Msun\ but these are ejected during the giant eruptions. However, one effect of the S Dor variations will be shaping of the circumstellar environment, potentially including the appearance of precursor or subsequent giant eruptions.

A small fraction of our numerically modeled stars do not satisfy the instability criteria - these are the lowest-mass and slowest-rotating stars in our sample. We expect such stars to evolve normally and {\it not} enter the S Dor variability phase that would mark them as LBVs. Our models of more massive or rapidly rotating stars do satisfy the instability criteria, suggesting that they will trigger S Dor variations during their lifetimes and thus could be observationally identified as LBVs.

Most of our models are unstable at both the equator and the poles, and it is worth considering the implications when combining the differences in \vescp\ and \vesceq\ with the possibility of a bistability jump between the poles and equator. If we assume that the ejection velocities scale with the escape speeds, and that for a star that has just arrived at the S Dor instability strip the mass lost via the instability mechanism will be ejected into a relatively homogenous medium around the star, we can take the ratio $\vinf({\rm p})/\vinf({\rm eq})$ as a velocity ``aspect ratio" that could correspond to the star's mass loss geometry.

In models where $\vinf({\rm p})/\vinf({\rm eq})$ is larger than 1 (the ratio is largest for models where \Tp\ is higher than the temperature of the bistability jump while \Teq\ is lower), this could manifest observationally as a wind geometry that appears elongated at the poles relative to the equator. By contrast, models where $\vinf({\rm p})/\vinf({\rm eq}) \sim 1$ might corresond to roughly spherical winds. Finally, if the instability is triggered only at the equator but not at the poles this may appear as a ``disk", while an instability triggered only at the poles would result in a ``bipolar" outflow. We include this ratio, and identify models where the instability is triggered at only the equator or the poles, in the last column of Table 3.

The correlation between rapid rotation, triggering the S Dor instability, and observed circumstellar geometry is certainly not an easy or exact one; as noted above, in rapidly rotating stars with high $L/M$, \geff\ will be more strongly reduced at the poles relative to the equator. However, we can see that for most of our models an increased $\omega$ corresponds to an increasingly likelihood that the star will hit the bistability jump and have a $\vinf({\rm p})/\vinf({\rm eq})$ much greater than 1.

This agrees with existing observations of rotation rates and circumstellar environments in known LBVs and S Dor variables, as well as candidate members of both classes:

{\it AG Car, HR Car, and R 127:} The Galactic LBV and S Dor variable AG Car has a $\vrot=190$ km s$^{-1}$ (corresponding to $\omega \sim 0.86$) and a bipolar nebula (see \citealt{Groh06,Groh09,WeisBomans20}). Another Galactic LBV and S Dor variable, HR Car, has $\vrot=190$ km s$^{-1}$ ($\omega \sim 0.88$) and a well-studied bipolar nebula with a known inclination angle \citep{Davies05,Groh09}. It is important to note that our results indicate that these stars will also produce an equatorial disk of circumstellar material. Unfortunately, most LBV nebulae have not been imaged in enough detail, or at close enough inner working angles, to distinguish such a feature. However, spectropolarimetry of AG Car has suggested the presence of both polar- {\it and} equatorially-driven winds \citep{Schulte-Ladbeck94,Davies05}. \cite{WeisBomans20} note that a significant fraction of known and candidate LBVs show some degree of bipolarity, ranging from strong hourglass shapes (including AG Car and HR Car) to weaker bipolar attachments (including known S Dor variable R 127).

{\it HD 160529}: By contrast, the Galactic LBV and S Dor variable HD 160529 lacks a prominent or asymmetric circumstellar nebula (e.g. \citealt{Smith02}), and observations of this star suggest a relatively low rotation rate of $v$ $sin$ $i\le45$ km s$^{-1}$ \citep{Stahl03}. This agrees with the implication from our numerical models that LBVs with a middling initial mass and low $\omega$ will display S Dor variability along with roughly circular circumstellar morphologies. \cite{WeisBomans20} identify a number of other well-studied LBVs or LBV candidates with simple spherical morphologies, including He3-519, S61, and P Cygni (for the latter see \citealt{Gootkin20} as well).

{\it SN 2009ip and R127}: We noted above that several of our models will hit the instability limit only at equatorial latitudes, implying the presence of a disk- or ring-like circumstellar structure. This is also supported by observations. Observations of the LMC LBV and S Dor variable R127 suggest that this star may have an equatorially-enhanced wind with minimal contributions from bipolar mass loss \citep{Schulte-Ladbeck93,Davies05}. In addition, the LBV candidate SN 2009ip was initially discovered during a mass outburst and eventually classified as an LBV before undergoing what appears to have been a terminal Type IIn supernova in 2012 (e.g. \citealt{Margutti14,Graham17}). \cite{Levesque14} found that a supernova shock colliding with a pre-existing thin circumstellar disk at the star's equator was the best model to explain the evolving spectroscopy of the 2012 supernova, suggesting that the disk had been produced by previous variations of the LBV. {\cite{Reilly17}} obtained detailed polarimetry of the SN 2009ip system and arrived at a similar model identifying the presence of an equatorial circumstellar disk that predated the 2012 supernova.  Both of these examples have also evolved in low-metallicity environments, in agreement with our results that find circumstellar equatorial disk geometries in the $Z=0.002$ stellar models only.

It is, of course, difficult to carry out reliable one-to-one comparisons between our numerical models and the handful of confirmed S Dor variables with well-determined physical properties, rotation rates, and circumstellar environments. However, it is encouraging to note the many similarities between our models' predictions and available observational data.


  \section{Summary, discussion and conclusions} \label{sec:conclusions}

We have studied the triggering of S Dor instabilities in rotating massive post-main stars in the LBV instability strip. Adopting the UF98 criterion for atmosphere stability, we calculated equatorial and polar stability for a grid of stellar models with varying values of $L/M$, $X_H$, and $\omega$, based on the Geneva models for solar and sub-solar metallicity stars with mass loss and initial rotation rates of $\omi=0$ and 0.4 \citep{Ekstrom12,Georgy13} for massive stars close to the S Dor strip. The results of these models are listed in Table 3 and shown in Figs \ref{fig:Gal} and \ref{fig:SMC}.

In this study of the triggering of the S Dor variations in rotating and non-rotating stars, we have made some simplifying assumptions. Testing the validity of these assumptions will require full 3-D hydrodynamical models.

(a) We have assumed that the instability is triggered when the effective surface gravity is reduced by rotation and radiation pressure to only 10 \%\ of the Newtonian gravity. This assumption is based on the work of \citet{UF98} who showed that the corresponding predicted location of massive stars in the HRD agrees with the observed location of LBVs in their hot phase. The low effective gravity implies 
a high mass-loss rate, which is a requirement for the occurence of S Dor variations (see Introduction).

(b) For calculating the effective gravity of rotating stars, we assumed that the angular rotation rate $\Omega$ is independent of latitude  (see Eq. (9)). This 
implies that we ignore possible deviations due to differential rotation as a result of initial conditions or induced by magnetic fields.

(c) Our predictions are made for massive luminous stars with $\log(L/\Lsun) > 5.5$ ($\ge$32~M$_{\odot}$), i.e. for stars that are not expected to become RSGs (see also \citealt{Massey23}).
This implies that we do not consider lower luminosity hot massive post-RSG stars, although they
may show similar S Dor variations when their effective gravity ratio is reduced to $\geff/\gN = 0.10$.

Below we summarize several key takeaways from this work and note the observational efforts that would be most valuable for testing this proposed explanation for how S Dor variations are triggered in LBVs. In particular, we propose that 1) determining rotation rates for as many LBVs as possible, 2) a census of LBV prevalence as a function of metallicity, and 3) a large high-contrast imaging survey of LBV circumstellar environments would be invaluable for testing the results presented here:

\begin{enumerate}
\item{During post-MS evolution, when massive stars expand with \Teff\ decreasing at almost constant $L /M$, the equators of rotating stars become unstable earlier (i.e. at higher values of \Teff) than those of non-rotating stars. This is because rotation lowers the equatorial temperature and increases \kapmax. The faster the rotation rate, the earlier the instability sets in. A range of rotation rates will thus broaden the observed LBV strip in the HRD (see Figure \ref{fig:fig1}). Given the importance of rotation in triggering  the S Dor instability, and the expected impact of rotation rates on the observed population of LBVs, a comprehensive observational survey of LBV rotation rates is urgently needed.}

\item{We found that Galactic models with $\Mi > 40 \Msun$ and the values of $L$ and $\Teff$ adopted from the location of the LBV instability strip (Figure \ref{fig:fig1}) are all unstable at the equator and the poles, even without rotation. For Galactic stars of  $\Mi \le 40 \Msun$ the situation is more complex. If we adopt the $L/M$ ratios derived from evolution models without rotation, the models are more stable than those with $L/M$ adopted from evolution models with rotation because the latter have a higher luminosity and a lower mass (Figure \ref{fig:Gal} and 13).}

\item{SMC models are more stable than Galactic models for two reasons: they have comparable luminosities but higher masses than their Galactic counterparts in the post-MS phase due to their lower mass-loss rates, resulting in smaller values of $L/M$, and the opacity is smaller in the  SMC stars because of their lower metallicity. The difference between the predicted stability of Galactic and SMC models can be seen in Figure 9 and in the comparison between Figures \ref{fig:Gal} and \ref{fig:SMC}. With this result in mind, an observational survey of LBV population as a function of metallicity would be extremely valuable, though we recognize that a complete search for candidates and confirmation of their nature as LBVs is a complex undertaking (e.g. \citealt{RichardsonMehner18}, \citealt{WeisBomans20}).}

\item{The equators of rotating stars become unstable before the polar regions, but the implications of this effect are complex. At small values of $\omega \le 0.3$ the ratio between polar and equatorial temperature is $<1.05$ (Eq. \ref{eq:Tratio}) and the ratio in $\kapmax$ is $>0.96$. As a result, the polar regions are only {\it marginally} stable when the equator becomes unstable. In that case the instability at the equator will trigger an instability over the full surface of the star. On the other hand, at high rotation rates the polar regions are still stable when the equator becomes unstable.  We expect that in that case the launching of the equatorial wind may result in a pole-to-equator flow along the surface of the star and strong equatorial mass loss.  Evidence of equatorial rings has been observed for LBVs such as SN 2009ip and R127; with a larger observational sample of LBV rotation rates and circumstellar geometries, it would be interesting to test whether there is a correlation between the rotation rate of LBVs and the presence of equatorial disks.}

\item{The terminal velocities of the radiation-driven winds from the equatorial and polar regions are expected to scale with the local escape velocity  (e.g. \citealt{Kudritzki89,LamersSnowLindholm95,LamersCassinelli99}). If both the equator and poles are unstable, a dense low-velocity wind will be launched from the equator and a faster wind will be launched from the poles. As a result, the material ejected during the S Dor variations of rapidly rotating LBVs is expected to be prolate, with a lower density and higher velocity in the polar regions. While the circumstellar geometry of LBVs is expected to be complex, shaped by S Dor variations as well as instabilities on faster \citep{Jiang18} and slower (giant eruption) timescales, these results suggest a prolate appearance with a lower density and higher velocity in the polar regions. This agrees with observations of rotation rates and circumstellar nebulae for LBVs such as AG Car and HR Car; however, the relationship between rotation rate and circumstellar geometry is further complicated by factors such as metallicity effects (for example, Table 2 shows that massive stars with lower metallicities are able to maintain rapid rotation during their main sequence evolution due to lower mass-loss rates) and binary interactions. A large imaging sample of LBV circumstellar environments is needed in order to fully understand the phenomena and timescale shaping their geometries.}

\item{We have adopted the proximity of the Modified Eddington Limit as a condition for triggering the S Dor variations, because UF98 showed that this explains the location of the hot side of the LBV strip in the HRD.
However, the model by \cite{grassitelli21} suggests that the S Dor variability is due to changes in mass-loss rate, resulting in temperature and opacity variations in the {\it subphotospheric} extended envelope. The relation between these two aspects can be understood because a very small effective gravity in the photosphere will result in a high mass-loss rate, which is a condition for the triggering of the S Dor variations in the model by \cite{grassitelli21}. }

\item{LBVs are located in a part of the HRD that is also occupied by stars that do not show (or have not yet shown) S Dor variability. This is not surprising considering the critical conditions that we have derived for triggering the  variations, in particular the critical $L/M$ ratio and the rotation rate. 
Firstly, stars may enter the post-main sequence phase with a broad range of rotation rates, and some of them may reach the critical rotation-dependent MEL later than others or not at all. Second, the triggering criterium depends sensitively on the L/M ratio (see Fig. 9 for non-rotating stars and Fig 10 for rotating stars) which in turn depends on their previous evolution. 
Evolution calculations show that massive stars have to lose a considerable fraction of their mass in the LBV phase to explain the absence of very luminous RSGs above the Humphreys-Davidson limit \citep{HumphreysDavidson79,HumphreysDvidson94}. This mass may be lost in stellar winds, which is sensitive to rotation, in the S Dor phase, or even in giant eruptions (\citealt{Lamers89,Davidson87,Davidson20}). All these effects will determine whether, and if so, at what mass and luminosity, the star may reach the critical conditions for triggering LBV variations. The lower the initial mass of the stars, the less likely it is that they reach the critical L/M ratio and the larger the fraction of normal stars in the S Dor instability strip (see also point 5 in section 5.3).}

\item{Finally, we note that our semi-analytical models for the triggering of S Dor variations of rotating star in the LBV instability strip can be improved by future detailed hydrodynamical simulations. The main improvements to be expected from these models include (a) calculating the stability of photospheres of stars with non-LTE model atmospheres and consistent evolution models of rotating stars, (b) calculating propagation of the photospheric instability into deeper layers to explain the expansion, the mass-loss rate, and the duration of the S Dor variations (a result that could in turn have useful implications for testing these models with time-domain observations of LBVs), and (c) calculating the effect of the geometry of the mass ejected during the S Dor variations of rotating stars on the shaping of circumstellar material ejected during other mass loss events such as giant eruptions.}
\end{enumerate}

\acknowledgements 
 We are grateful for the contributions of our anonymous referee, whose detailed feedback was instrumental in improving the final version of the paper. We thank Jose Groh, Norbert Langer, Jamie Lomax, Andre Maeder, and Philip Massey for very valuable conversations. This research was supported by NSF grant AST 1714285, an Alfred P. Sloan fellowship, and a Guggenheim fellowship awarded to EML. HJGLM is grateful for hospitality at the University of Washington and for travel grants from the Leidsch Kerkhoven-Bosscha Fonds, NOVA and the Astronomical Institute Anton Pannekoek of the University of Amsterdam.

  
\appendix
\section{Power Law Fits}
\label{sec:powerlawfits}

This Appendix describes some of the power-law approximations that we have used
to study the triggering of the instability in rotating stars.
\\

\subsection{The function \fT} 

The dependence of \kapmax\ on \Teff\ is given by
$ \kapmax = 0.313 / \fT$    
see Eq. \ref{eq:kapmaxf}. The function \fT\ can be approximated 
with an accuracy of 0.01 by

  \begin{equation} 
  \begin{split}
  \fT (Z=0.02) = & 1.205-5.002 x +11.466 x^2  \\
                & -11.905 x^3 +4.685 x^4 \\
  \fT (Z=0.002)= & 1.289 -4.373 x +9.099 x^2 \\
                 & -8.822 x^3 +3.357 x^4 \\
  \end{split}
  \label{eq:App-fT}
  \end{equation}
  with $x=10^4/\Teff$.
  \\

\subsection{The function $\Psi(\omega)$}

The function $\Psi(\omega)=L/\Lp$, defined by Eq. \ref{eq:Psi}, describes the ratio between the total 
luminosity and the polar luminosity of rotating stars.  It can be approximated with a accuracy of 0.007 by
   
  \begin{equation} 
  \Psi(\omega) =  1.000 +0.0936 ~ \omega -1.0887 ~ \omega^2 +0.5956 ~ \omega^3
  \label{eq:AppPsi}
  \end{equation}
\\

\subsection{The function $R(\Theta)/\Rp$} 

The shape of a rotating star is given by the solution of Eq. \ref{eq:shapefit}. The solution 
can be written as $x(y)$ where $x\equiv \Rtheta/\Rp$ and $y=B \sin^2(\Theta)$ with
$B=3.75 \omega^2 / (1+0.5 \omega^2)^3 $. Notice that $B=1$ if $\omega = 1$ and so $0<y<1$ and $1 < x < 1.5$.
The function $x(y)$ can be described by a power law approximation

\begin{equation}
 \begin{split}
 R(\Theta)/\Rp \equiv &  x(y) =  1.000+0.165y + 0.125 y^3 \\
               & + 0.125y^{11}+0.085 y^{100}  \\
 \end{split}
  \label{eq:RthetaRp}
\end{equation}

with an accuracy of $\Delta x < 0.002$ at $ 0 < y< 0.98$ and $\Delta x < 0.02$ at $ 0.98 < y< 1.00$.

\subsection{The function $s(\omega)$}  

The surface $S(\Rp,\omega)= 4 \pi \Rp^2 s(\omega)$ of a rotating star with polar radius \Rp\ and rotation rate 
$\omega$, described by Eq. \ref{eq:s}, can be approximated with an accuracy of 0.0006 by the power law approximation
 
 \begin{equation} 
 s(\omega) \simeq 1.00+0.0254\omega+ 0.2274\omega^2 + 0.2040 \omega^3 - 0.2467 \omega^4
 \label{eq:sapprox}
 \end{equation}
 with $1 < s(\omega) < 1.21$. 
 
\bibliography{LevesqueLamers}
\bibliographystyle{aasjournal}

\clearpage
\startlongtable
\begin{longrotatetable}
\begin{deluxetable*}{c c c c c c c c c c c c c c c c}
\tablewidth{0pt}
\tabletypesize{\footnotesize}
\tablecolumns{15}
\tablecaption{Parameters of rotating stars in the S Dor instability strip}
\tablehead{\colhead{M$_i$} & \colhead{M} & \colhead{log(L)} & \colhead{log(L/M)} & \colhead{log(\Tmean)} & \colhead{$X_H$} 
          & \colhead{$\omega$} & \colhead{\Rp} & \colhead{$\Req$} & \colhead{log(\Tp)} & \colhead{log(\Teq)}    
          & \colhead{$\vroteq$} & \colhead{$\vescp$} & \colhead{$\vesceq$}
          & \colhead{Instab?} & \colhead{$\vinf({\rm p})/\vinf({\rm eq})$} \\
\colhead{(\Msun)} & \colhead{(\Msun)} & \colhead{(\Lsun)} & \colhead{(\Lsun/\Msun)} & \colhead{K} & \colhead{} & \colhead{} 
         & \colhead{(\Rsun)} & \colhead{(\Rsun)} & \colhead{(K)} & \colhead{(K)}
         & \colhead{(km/s)}   & \colhead{(km/s)} & \colhead{(km/s)} & \colhead{} & \colhead{}
         }
\startdata
\hline
\hline
{\bf Z=0.014} & {\bf no rot} & & & & & & & & & & & & & & \\
\hline
 32  & 30.1  & 5.50 &4.021 & 4.129  & 0.72  & 0.0  & 103.4  & 103.4  & 4.129  & 4.129  &0 &  284  &  284  & N   &  \nodata \\
  & & & & & & 0.3  & 101.8  & 106.4  & 4.138  & 4.119  &   60  &  284  &  269  & N    &  \nodata \\
  & & & & & & 0.6  &  97.5  & 115.0  & 4.167  & 4.083  &  119  &  278  &  224  & N    &  \nodata \\
  & & & & & & 0.9  &  92.6  & 130.1  & 4.202  & 3.948  &  182  &  266  &  124  & Y$_p$   &  bipolar \\
 40  & 36.5  & 5.67 &4.108  & 4.237  & 0.72  & 0.0  &  76.5  &  76.5  & 4.237  & 4.237  &0.0  & 349  &  349  & N    & \nodata \\
  & & & & & & 0.3  &  75.3  &  78.7  & 4.246  & 4.227  &   74  &  346  &  331  & N    &  \nodata \\
  & & & & & & 0.6  &  72.1  &  85.1  & 4.275  & 4.191  &  147  &  335  &  277  & Y    &  1.21 \\
  & & & & & & 0.9  &  68.5  &  96.2  & 4.310  & 4.056  &  230  &  311  &  158  & Y    &  1.97 \\
 60  & 36.2  & 5.91 &4.351  & 4.343  & 0.49  & 0.0  &  61.9  &  61.9  & 4.343  & 4.343  &0.0  & 332  &  332  & Y    &  1.00 \\
  & & & & & & 0.3  &  60.9  &  63.7  & 4.352  & 4.333  &   71  &  324  &  318  & Y    &  1.02 \\
  & & & & & & 0.6  &  58.3  &  68.8  & 4.381  & 4.297  &  146  &  292  &  276 & Y    &  2.12 \\
  & & & & & & 0.9  &  55.4  &  77.9  & 4.416  & 4.162  &  247  &  223  &  1689 & Y    &  2.64 \\
 85  & 49.2  & 6.15 &4.458  & 4.455  & 0.36  & 0.0  &  48.7  &  48.7  & 4.455  & 4.455  &0.0  & 396  &  396  & Y    &  1.00 \\
  & & & & & & 0.3  &  48.0  &  50.1  & 4.464  & 4.445  &   85  &  382  &  383  & Y    &  1.00 \\
  & & & & & & 0.6  &  45.9  &  54.2  & 4.493  & 4.409  &  181  &  320  &  340  & Y    &  0.94 \\
  & & & & & & 0.9  &  43.6  &  61.3  & 4.528  & 4.274  &  320  &  168  &  219  & Y    &  1.53 \\
\hline
{\bf Z=0.014} & {\bf rot   } & & & & & & & & & & & & & & \\
\hline
 32  & 28.2  & 5.64 &4.190  & 4.239  & 0.54  & 0.0  &  73.2  &  73.2  & 4.239  & 4.239  &0  & 307  &  307  & Y   &  1.00 \\
  & & & & & & 0.3  &  72.1  &  75.3  & 4.248  & 4.229  &   65  &  304  &  292  & Y    & 1.04 \\
  & & & & & & 0.6  &  69.0  &  81.4  & 4.277  & 4.193  &  130  &  291  &  245  & Y    & 1.19 \\
  & & & & & & 0.9  &  65.6  &  92.1  & 4.312  & 4.058  &  206  &  266  &  141  & Y    & 1.89 \\
 40  & 32.0  & 5.82 &4.315  & 4.314  & 0.43  & 0.0  &  63.8  &  63.8  & 4.314  & 4.314  &0.0  & 325  &  325  & Y    &  1.00 \\
  & & & & & & 0.3  &  62.8  &  65.6  & 4.323  & 4.304  &   69  &  320  &  311  & Y    &  2.06 \\
  & & & & & & 0.6  &  60.1  &  70.9  & 4.352  & 4.268  &  141  &  297  &  266  & Y    &  2.24 \\
  & & & & & & 0.9  &  57.1  &  80.2  & 4.387  & 4.133  &  232  &  251  &  159  & Y    &  3.17 \\
 60  & 38.3  & 6.12 &4.537  & 4.436  & 0.17  & 0.0  &  51.4  &  51.4  & 4.436  & 4.436  &0.0  & 332  &  332  & Y    &  1.00 \\ 
  & & & & & & 0.3  &  50.6  &  52.8  & 4.445  & 4.426  &   72  &  320  &  322  & Y    &  0.99 \\ 
  & & & & & & 0.6  &  48.4  &  57.1  & 4.474  & 4.390  &  153  &  262  &  288  & Y    &  0.91 \\
  & & & & & & 0.9  &  46.0  &  64.6  & 4.509  & 4.255  &  274  &  107  &  189  & Y    &  1.14 \\
\hline
{\bf Z=0.002} & {\bf no rot} & & & & & & & & & & & & & & \\
\hline
 32  & 31.3  & 5.59 &4.094  & 4.210  & 0.75  & 0.0  &  79.0  &  79.0  & 4.210  & 4.210  &0  &  319  &  319  & N    & \nodata \\
  & & & & & & 0.3  &  77.8  &  81.3  & 4.219  & 4.200  &   67  &  317  &  303  & N    &  \nodata \\
  & & & & & & 0.6  &  74.5  &  87.9  & 4.248  & 4.164  &  134  &  306  &  253  & N    &  \nodata \\
  & & & & & & 0.9  &  70.7  & 99.4  & 4.283  & 4.029  &  210  &  285  &  144  & Y    &  1.98 \\
 40  & 39.1  & 5.75 &4.158  & 4.331  & 0.75  & 0.0  &  54.4  &  54.4  & 4.331  & 4.331  &0.0  &  412  & 412 & N    & \nodata \\
  & & & & & & 0.3  &  53.6  &  56.0  & 4.340  & 4.321  &   87  &  408  &  392  & N    &  \nodata \\
  & & & & & & 0.6  &  51.3  &  60.5  & 4.369  & 4.285  &  176  &  389  &  331  & Y$_{eq}$  &  disk \\
  & & & & & & 0.9  &  48.7  &  68.5  & 4.404  & 4.150  &  280  &  350  &  192  & Y    &  3.64 \\
 60  & 57.8  & 6.02 &4.258  & 4.393  & 0.75  & 0.0  &  55.8  &  55.8  & 4.393  & 4.393  &0.0  & 453  &  453  & N    & \nodata \\
  & & & & & & 0.3  &  54.9  &  57.4  & 4.402  & 4.383  &   96  &  445  &  434  & Y$_{eq}$  & disk\\
  & & & & & & 0.6  &  52.6  &  62.1  & 4.431  & 4.347  &  198  &  406  &  373  & Y    &  1.09 \\
  & & & & & & 0.9  &  50.0  &  70.2  & 4.466  & 4.212  &  331  &  327  &  227  & Y    & 2.89 \\
 85  & 80.9  & 6.23 &4.322  & 4.479  & 0.75  & 0.0  &  47.8  &  47.8  & 4.479  & 4.479  &0  & 535  &  535  & Y & 1.00 \\
  & & & & & & 0.3  &  47.1  &  49.2  & 4.488  & 4.469  &  114  &  519  &  515  & Y    &  1.01 \\
  & & & & & & 0.6  &  45.1  &  53.2  & 4.517  & 4.433  &  240  &  450  &  452  & Y    &  0.99 \\
  & & & & & & 0.9  &  42.8  &  60.2  & 4.552  & 4.298  &  417  &  296  &  286  & Y    &  2.07 \\
\hline
{\bf Z=0.002} & {\bf rot   } & & & & & & & & & & & & & & \\
\hline
 32  & 31.2  & 5.65 &4.156  & 4.225  & 0.69  & 0.0  &  79.0  &  79.0  & 4.225  & 4.225  &0  & 309  &  309  & N    & \nodata \\
  & & & & & & 0.3  &  77.8  &  81.3  & 4.234  & 4.215 &   65  &   307  &  294  & N    &  \nodata \\
  & & & & & & 0.6  &  74.5  &  87.9  & 4.263  & 4.179  &  131  &  293  &  247  & Y$_{eq}$    & disk \\
  & & & & & & 0.9  &  70.7  & 99.4  & 4.298  & 4.044  &  208   &  267  &  143  & Y    &  1.87 \\
 40  & 38.7  & 5.80 &4.212  & 4.303  & 0.68  & 0.0  &  65.6  &  65.6  & 4.303  & 4.303  &0.0  & 363  &  363  & N    & \nodata \\
  & & & & & & 0.3  &  64.5  &  67.4  & 4.312  & 4.293  &   77  &  359  &  346  & N    & \nodata \\
  & & & & & & 0.6  &  61.8  &  72.9  & 4.341  & 4.257  &  156  &  337  &  294  & Y    & 2.30 \\
  & & & & & & 0.9  &  58.7  &  82.5  & 4.376  & 4.122  &  253  &  295  &  173  & Y    & 3.41 \\
 60  & 56.3  & 6.11 &4.359  & 4.426  & 0.60  & 0.0  &  53.1  &  53.1  & 4.426  & 4.426  &0.0  &  424  &  424  & Y    & 1.00 \\
  & & & & & & 0.3  &  52.3  &  54.7  & 4.435  & 4.416  &   91  &  412  &  408  & Y    & 1.01 \\
  & & & & & & 0.6  &  50.1  &  59.1  & 4.464  & 4.380  &  190  &  357  &  358  & Y    & 1.00 \\
  & & & & & & 0.9  &  47.6  &  66.9  & 4.499  & 4.245  &  330  &  237  &  226  & Y    &  2.09 \\
 85  & 77.2  & 6.36 &4.472  & 4.540  & 0.46  & 0.0  &  41.9  &  41.9  & 4.540  & 4.540  &0.0  &  490  &  490  & Y    &  1.00 \\
  & & & & & & 0.3  &  41.3  &  43.1  & 4.549  & 4.530  &  106  &  466  &  478  & Y    &  0.98 \\
  & & & & & & 0.6  &  39.5  &  46.7  & 4.578  & 4.494  &  231  &  354  &  436  & Y    &  0.81 \\
\hline
\enddata
\label{tbl:models}
\end{deluxetable*}
\end{longrotatetable}
\clearpage
\end{document}